\documentclass[%
 aip,
 amsmath,amssymb,
 reprint,%
]{revtex4-1}
\usepackage{xcolor}
\usepackage{graphicx}
\usepackage{float}
\usepackage{dcolumn}
\usepackage{bm}

\usepackage[utf8]{inputenc}
\usepackage[T1]{fontenc}
\usepackage{mathptmx}
\usepackage{etoolbox}

\usepackage{enumitem}
\setlist[enumerate]{itemsep=0mm}
\setlist[itemize]{itemsep=0mm} 
\makeatletter

\def\@email#1#2{%
 \endgroup
 \patchcmd{\titleblock@produce}
  {\frontmatter@RRAPformat}
  {\frontmatter@RRAPformat{\produce@RRAP{*#1\href{mailto:#2}{#2}}}\frontmatter@RRAPformat}
  {}{}
}%
\makeatother
\begin{document}

\preprint{AIP/123-QED}

\title{Structural Origins and Real-Time Drivers of Intermittency}

\author{A.~Barone}
\affiliation{%
 Department of Physics and Astronomy, University of Bologna, Viale Carlo Berti Pichat, 6/2, Bologna 40127, Italy
}%

\email{alessandro.barone7@unibo.it}

\author{A.~Carrassi}
\affiliation{%
Department of Physics and Astronomy, University of Bologna, Viale Carlo Berti Pichat, 6/2, Bologna 40127, Italy
}

\author{T.~Savary}
\affiliation{%
Royal Meteorological Institute of Belgium, Meteorological and Climatological Research, Avenue Circulaire, 3, 1180 Brussels, Belgium
}%

\author{J.~Demaeyer}
\affiliation{%
Royal Meteorological Institute of Belgium, Meteorological and Climatological Research, Avenue Circulaire, 3, 1180 Brussels, Belgium
}%

\author{S.~Vannitsem}
\affiliation{%
Royal Meteorological Institute of Belgium, Meteorological and Climatological Research, Avenue Circulaire, 3, 1180 Brussels, Belgium
}%

\date{25 June 2025}

\begin{abstract}

In general terms, intermittency is the property for which time evolving systems alternates among two or more different regimes. Predicting the instance when the regime switch will occur, is extremely challenging, often practically impossible. Intermittent processes include turbulence, convection, precipitation patterns, as well as several in plasma physics, medicine, neuroscience, and economics. Traditionally, focus has been on global statistical indicators, {\it e.g.} the average frequency of regime changes under fixed conditions, or how these vary as a function of the system’s parameters. We add a local perspective: we study the causes and drivers of the regime changes in real time, with the ultimate goal of predicting them. Using five different systems, of various complexities, we identify indicators and precursors of regime transitions that are common across the different intermittency mechanisms and dynamical models. For all the systems and intermittency types under study, we find a correlation between the alignment of some Lyapunov vectors and the concomitant, or aftermath, regime change. We discovered peculiar behaviors in the Lorenz 96 and in the Kuramoto-Shivanshinki models. In Lorenz 96 we identified crisis-induced intermittency with laminar intermissions, while in the Kuramoto-Shivanshinki we detected a spatially global intermittency which follows the scaling of type-I intermittency. The identification of general mechanisms driving intermittent behaviors, and in particular the unearthing of indicators spotting the regime change, pave the way to designing prediction tools in more realistic scenarios. These include turbulent geophysical fluids, rainfall patterns, or atmospheric deep convection. 

\end{abstract}

\maketitle

\begin{quotation}
Predicting regime changes in systems that alternate between different states in an apparently random manner is of primary importance to prevent catastrophic outcomes and guide mitigation measures. Intermittent dynamics, with its remarkable variety, represents an excellent prototype for this type of behavior. Research in the field of dynamical systems on intermittency has so far primarily focused on global/statistical analyzing the systems' structural changes that lead to the emergence of the intermittency. Estimating the average return time of a ``dangerous'' regime or studying the systems' responses to varying parameters or external forcing are notable examples from the realm of global analysis.  

In this study, we attempt moving beyond this and introduce a local perspective. Here we also look for a class of spatio-temporal local indicators for understanding and in the best case predicting regime changes. We use tools belonging to the Lyapunov theory, such as the covariant Lyapunov vectors (CLVs), and apply our analysis to five different dynamical systems.  

These methods can be applied to geophysical systems, and their generality may potentially allow for applications to other systems operating across different spatial and temporal scales. 
\end{quotation}

\section{\label{sec:Intro} Introduction}

Among the ``routes to chaos'', one prominent example is the intermittency pathway. For long, this process has been understood exclusively as the transition from periodic, or laminar, dynamics to chaotic behaviors. Over time, our understanding of intermittency has evolved to encompass its diverse manifestations in different contexts. 

Given the ubiquity of chaotic phenomena, it is possible to find intermittency in many and diverse systems, including forced nonlinear oscillators, electronic circuits, plasma physics, economics, medicine, or neuroscience \cite{elaskar_new_2017}. The term intermittency originally emerged in the context of turbulent flows to describe velocity fluctuations \cite{turb_flows_1949,elaskar_new_2017}. Intermittency is widely observed in turbulent atmospheric flows \cite{turb_mahrt_intermittent_1989,Lovejoy_Schertzer_2013, turb_zorzetto_extremes_2018,turb_liu_atmospheric_2023}, in the presence of convection \cite{conv_zumbrunnen_convective_1993,conv_ching_intermittency_2000,conv_garcia_bursting_2003,conv_chowdhuri_visibility_2021}, in rainfall data \cite{rain_schmitt_modeling_1998,rain_hess-17-355-2013, rain_schleiss_how_2018}, but also in the sea ice \cite{tc-13-2457-2019,amslaurea30307}. Recent studies have highlighted the intermittent nature of some phenomena on climatological scales \cite{clim_molini_revisiting_2009, clim_manucharyan_climate_2011, clim_giannakis_comparing_2012,clim_bellucci_intermittent_2022} and how climate change could affect key features of observed intermittencies \cite{change_schleiss_how_2018, change_sauquet_predicting_2021, change_moidu_spatial_2021}.

The study of intermittency has traditionally focused on understanding its global characteristics: {\it e.g.}, how frequently, on average, one should expect regime changes or how the frequency of these regime changes varies when certain structural properties of the system are altered, {\it e.g.}, model parameters or external forcing. Intermittency can thus emerge as an ``exceptional'' behavior, for a specific, typically small range of parameters' values, between systematic different behaviors associated with larger ranges of parameters.

In this work, we leverage this global analysis and unveil the specific intermittency types and mechanisms in a set of prototypical dynamical systems. 

We then move beyond and perform a (spatio-temporal) ``local'' analysis: we identify local indicators and, in the best cases, precursors of regime change in real time, across different types of intermittency.

The search of local indicators is performed through the lens of dynamical systems theory, with a focus on Lyapunov exponents and covariant Lyapunov vectors (CLVs). Recent studies have demonstrated that the alignment of certain CLVs can serve as predictors for the onset of large-amplitude peaks or regime transitions in specific dynamical systems \cite{CriticalTran, CLV_Alignment, viennet_guidelines_2022,brugnago_machine_2020} (see Sec.~\ref{sec:CLVs}). Building on this, we investigate whether CLVs are reliable indicators of regime transitions in a wider class of intermittent systems. We show that CLVs carry unique information that correlate with the regime transition. This opens up avenues of conjecture on the dynamical mechanisms underlying intermittency but also, in certain cases, means of predicting transitions. We will do that in a set of prototypical low-to-medium-dimensional dynamical systems, exemplars of different types of intermittency. This will shed new light on, and in some cases we will discover for the first time, the intermittency in well-known dynamical systems, including Lorenz 63 \cite{lorenz_deterministic_1963} and 96 \cite{Lorenz_2006} models and Kuramoto-Shivashinki \cite{Kuramoto1976PersistentPO, SIVASHINSKY19771177}.

The paper is structured as follows: in Sec.~\ref{sec:Int} we review classical intermittency theory and describe the different types of intermittency. In Sec.~\ref{sec:CLVs} we briefly introduce how we shall use Lyapunov tools to perform the global and local analysis. Section~\ref{sec:Results} presents the numerical results in five different systems. Final conclusions and perspective are drawn in Sec.~\ref{sec:concl}.

\begin{table*}
\caption{\label{tab:bifurcation_table}Summary of intermittencies and their characteristics. Columns indicate: \textbf{Type} — intermittency classification; \textbf{Description} — underlying mechanism; \textbf{Multipliers} — characteristic linear stability indicators; $\mathbf{\overline{l}}$ — mean laminar phase duration; \textbf{Char. relation} — scaling law near onset. For \textit{crisis-induced} intermittency, the mean burst time $\overline{\tau}$ is used, as the dynamics alternate between distinct chaotic regimes without a true laminar phase.}
\begin{ruledtabular}
\begin{tabular}{c>{\centering\arraybackslash}p{4cm} >{\centering\arraybackslash}p{3cm} c >{\centering\arraybackslash}p{3cm}}
\textbf{Type} & \textbf{Description} & \textbf{Multipliers} & $\mathbf{\overline{l}}$ & \textbf{Char. relation} \\ \hline \\ 
Type-I   & Local map inverse tangent & $\mathcal{R}({\mu})$ at +1 & $\frac{1}{\sqrt{a \varepsilon}} \arctan\left(c \sqrt{\frac{a}{\varepsilon}}\right)$ & $\overline{l} \propto \epsilon^{-\frac{1}{2}}$ \\ \\
Type-II  & Local map Hopf bifurcation & Two conjugate $|\mu_{i}| >1$ simultaneously & $\frac{1}{c\sqrt{a \varepsilon}} \arctan \left(c \sqrt{\frac{a}{c}}\right)$ & $\overline{l} \propto \epsilon^{-\frac{1}{2}}$ \\ \\  
Type-III & Local map subcritical period-doubling bifurcation & $\mathcal{R}({\mu})$ at -1 & $\frac{1}{c\sqrt{a \varepsilon}} \arctan \left(c \sqrt{\frac{a}{c}}\right)$ & $\overline{l} \propto \epsilon^{-\frac{1}{2}}$ \\ \\ 
Type-V   & Stable fixed point in a nondifferentiable, even discontinuous map & Not trivial, depends on the cases & $\int_{x_{out}}^C b(x)P(x)dx + \frac{| \ln(\varepsilon)|}{a}$ & $\overline{l} \propto \ln(\varepsilon)$ \\ \\ 
Type-X   & Inverse tangent, always the same reinjection point & $\mathcal{R}({\mu})$ at +1  & $\frac{\ln(c/\rho)}{\mu}$ & $\overline{l}  \propto \ln(1/\varepsilon)$ \\ \\ 
On-Off   & Invariant object and orbits entering and leaving every neighborhood of it & &  & $\overline{l} \propto \frac{1}{\epsilon} $ \\ \\
Eyelet   & Sync. of unstable periodic saddle orbits embedded in chaotic attractors of coupled oscillators &  &  & $\overline{l}  \propto \exp\{\frac{cost}{\sqrt{\epsilon_c-\epsilon}}\}$ \\ \\  
Crisis-induced & Attractor widening or attractor merging crises &  & & $\overline{\tau }\propto |p-p_c|^{- \gamma}$\\ \\  
\end{tabular}
\end{ruledtabular}
\end{table*}

\section{\label{sec:Int} Intermittency in dynamical systems: A brief survey}

It is with the groundbreaking work of Pomeau and Manneville published in 1980 \cite{pomeau_intermittent_1980} that intermittency became fully integrated in the field of dynamical systems, providing us with an initial, yet ample, classification of this phenomenon. 
They identified three types of intermittency by analyzing the Lorenz system (L63) \cite{lorenz_deterministic_1963}, with the aid of Poincaré maps and the Floquet theory. In particular, using the Floquet multipliers, the Pomeau-Manneville (PM) classification distinguishes:
\begin{itemize}
    \item {\it Type-I intermittency}: A real eigenvalue crosses the unit circle at +1.
    \item {\it Type-II intermittency}: Two conjugate complex eigenvalues cross the unit circle simultaneously.
    \item {\it Type-III intermittency}: A real eigenvalue crosses the unit circle at $-1$.
\end{itemize}

The use of Floquet multipliers as an appropriate and powerful classifier for the first three types of intermittency is due to these types of intermittency being driven by the loss of stability of a periodic orbit via a bifurcation. Historically, such transitions have been analyzed through by studying the fixed points of Poincarè maps. The way these fixed points change their dynamical nature is precisely characterized by the Floquet multipliers, which therefore provide essential information about the global behavior of the system near the bifurcation threshold.

Two important dynamical-statistical parameters characterizing intermittency in general are the {\it laminar length}, $l$, and the {\it characteristic relation} \cite{elaskar_new_2017,nayfeh_applied_1995,schuster_deterministic_2005} (see Tab.~\ref{tab:bifurcation_table}). $l$ is the duration of the laminar phase, while the characteristic relation describes the scaling of the average laminar length with respect to the deviation of the model parameter from the critical value leading to intermittency. 

By leveraging simple dynamical system considerations, we can estimate $l$ and the characteristic relation for the case of Type-I intermittency. Similar arguments can be used for some other types of intermittency, specifically the ones that can be studied through the local Poincaré map ({\it i.e.}, intermittency driven by a bifurcation-like mechanism). Let us consider a nonlinear autonomous dynamical system with only one parameter, $p$,  

\begin{equation}
    \frac{\mathrm{d}\mathbf{x}}{\mathrm{d}t} = \mathbf{F}(\mathbf{x}, p),
    \label{eq:autonomo}
\end{equation}

where $\mathbf{x} \in \mathbb{R}^{{\rm N}}$, $\mathbf{F}: \mathbb{R}^{\rm N} \mapsto  \mathbb{R}^{\rm N}$, and $p \in \mathbb{R}$. Assume that the system displays type-I intermittency for some $p>p_c$, so that, according to the PM classification, a real Floquet multiplier leaves the unite circle at $+1$, \textit{i.e.} it moves away from the unit circle while staying aligned with the positive x-axis. We shall see that the manifold corresponding to this multiplier carries important information associated with intermittency, while in the other directions there is dissipation \cite{nayfeh_applied_1995, elaskar_new_2017}. We can use the local map associated with this manifold to describe the intermittent behavior of the system.  

The local map $x_{n+1}=F(x_n)$ depends on the deviation, $\varepsilon = |p-p_c|$, from the critical value $p_c$. For type-I intermittency, the local map undergoes the inverse tangent bifurcation, which reads:

\begin{equation}
    x_{n+1} = \varepsilon + x_n + a x_n^2.
\label{eq:interI}
\end{equation}

As long as $a \ne0$, the quadratic term does not alter the qualitative behavior of the map, but its amplitude changes the locations of its fixed points. Their existence is determined by $\varepsilon$, independently from $a$ (for $a \ne0$). For $\varepsilon < 0$ the map presents two fixed points, one stable and one unstable. At $\varepsilon = 0$ they merge into one and the Floquet multiplier becomes +1. For $\varepsilon > 0$ there are not fixed points. This means that the local map goes through an inverse tangent bifurcation. 

The map transition through the bifurcation can be further analyzed with the aid of the Cobweb plot (not shown here for brevity). In the Cobweb plot, for $\varepsilon = |p-
p_c|  \gtrsim 0$, a narrow channel opens between the quadratic and the $x_{n+1}=x_n$ line. The solution gets trapped in this narrow channel within which the map takes many iterations before it finally escapes. The period of stay of the local map solution in this narrow channel corresponds to $l$.

When the trajectory escapes the laminar region, it marks the beginning of the chaotic phase, which lasts until the trajectory returns to the narrow channel and is trapped in it for the next laminar phase. This alternation is driven by the so-called re-injection mechanism \cite{nayfeh_applied_1995,schuster_deterministic_2005,elaskar_new_2017}, which in turn provides ways to estimate $l$. Given that there is not a specific re-injection point, the laminar phases last differently and their duration is treated probabilistically. 

The probability density function (PDF) of the trajectories being re-injected in a specific point $x$ is known as the re-injection probability density function (RPD), indicated here as $\phi(x)$. 
Obtaining relevant statistical estimators of the re-injection point, and thus of the duration of the laminar length, requires knowledge of, or strong assumption on,  $\phi(x)$. Usually, re-injection points are considered equiprobable, which corresponds to take $\phi(x)$ as a uniform distribution. However, recent studies \cite{elaskar_new_2017, new_TypeII, New_typeIII, new_type1}, have highlighted the possibility and the necessity of estimating the RPD on a case-by-case basis.

If we restrict our analysis to the channel close to the vanished fixed point ($\varepsilon = |p-p_c|  \gtrsim 0$), the difference between two successive iterations, $x_{n+1}-x_n$, becomes very small and can be approximated as $x_{n+1}-x_n \approx \mathrm{d}x/\mathrm{d}l$, so that: 

\begin{equation}
    \frac{\mathrm{dx}}{\mathrm{d}l} = {\varepsilon + ax^2} ,
\label{eq:I1}
\end{equation}

with $\mathrm{d}l$ being the differential of the laminar length. We can now integrate from the point of injection to the upper limit of the laminar interval in order to obtain the {\it laminar length}, $l$,

\begin{equation}
    {\int_{x}^{c} \frac{\mathrm{d}x}{\varepsilon + ax^2} = \int_{0}^{l} \mathrm{d}l}=l(x,c).
\label{eq:I2}
\end{equation}

The laminar length, $l(x,c)$, is obtained by integrating Eq.~\ref{eq:I2}, and is a function of x and of the upper limit $c$ ({\it i.e. }the maximum length of the laminar phase),

\begin{equation}
    l(x,c) = \frac{1}{\sqrt{a \varepsilon}}\left[arctan \left(\frac{c}{\sqrt{\varepsilon/a}}\right)-arctan\left(\frac{x}{\sqrt{\varepsilon/a}}\right)\right].
\label{eq:I3}
\end{equation}

Then, choosing the RPD as uniform, we can obtain the average laminar length as

\begin{equation}
   \overline{l} = \int_{-c}^{+c}\phi(x) l(x,c) \mathrm{d}l = 
    \frac{1}{\sqrt{a \varepsilon}} arctan\left(c \sqrt{\frac{a}{\varepsilon}}\right) , 
\label{eq:I7}
\end{equation}

with $-c$, being the lower limit of the laminar interval. Despite its theoretical power, Eq.~\ref{eq:I7} is of limited practical use unless one knows $c$, which is usually unknown. On the other hand, a useful scaling law can be obtained in certain limiting scenarios. In particular, in the vicinity of the bifurcation where $\epsilon~\gtrsim 0$, we have $c \sqrt{a / \varepsilon} >> 1$ and Eq.~\ref{eq:I7} reduces to 

\begin{equation}
    \overline{l} \propto \frac{1}{\sqrt{\varepsilon}}.
\label{eq:I8}
\end{equation}

Equation~\ref{eq:I8} is the characteristic relation \cite{pomeau_intermittent_1980}. The case $c \sqrt{a / \varepsilon} >> 1$, corresponds to very long laminar periods compared to the duration of the chaotic bursts. In general, the characteristic relation can be written as:
\begin{equation}
    \overline{l} \propto \varepsilon^{-\beta} ,
\label{eq:I9}
\end{equation}
where $\beta$ indicates the rate of change of the average laminar length $\overline{l}$ as a function of the parameter $\varepsilon$. 

Going beyond the PM classification, other forms of intermittency exist. A summary of them, together with their main characteristics, is given in Table~\ref{tab:bifurcation_table}. 

Type-V intermittency appears when a stable fixed point of a map collides with a non-differentiable or discontinuous point of this map~\cite{nasiri_type-v_2022,elaskar_new_2017}. This type of intermittency has been named after the shape of the local map, where the V represent the vertices of the map around the non-differentiable or discontinuous point. In contrast to what we observe in the Cobweb plot for type-I, in this case it is not possible to identify the point where the map is tangent to the line $x_{n+1} = x_n$, since this point is non-differentiable or discontinuous.

Type-X intermittency~\cite{PRICE199129} has the same local map of type-I, but is characterized by a hysteresis process and a constant re-injection mechanism where the re-injected point is always the same ({\it i.e.} the RPD is a Dirac delta). 

In the on-off intermittency~\cite{platt_-off_1993, On-off2}, the system's solution jumps from being almost constant (the ``off'' phase) to activity (the ``on''  phase) and then back. On-off intermittency has two ingredients: (1) an invariant object, and, (2) orbits entering and leaving every neighborhood of the invariant object~\cite{elaskar_new_2017}. It can be modeled as:

\begin{equation}
\begin{cases}
  \frac{\mathrm{d} \mathbf{X}(t)}{\mathrm{dt}} = \mathbf{f}_1(\mathbf{X}(t), p (\mathbf{Y}(t))) \\
  \frac{\mathrm{d} \mathbf{Y}(t)}{\mathrm{dt}} = \mathbf{f}_2(\mathbf{Y}(t)) ,
\end{cases}
\label{eq:platt_sysa}
\end{equation}

whereas $\mathbf{X}(t)$ depends on $\mathbf{Y}(t)$ through a scalar control function $p : \mathbb{R}^{S-N} \to \mathbb{R}$. Here, $S$ denotes the dimension of the phase space, and $N$ is the dimension of the hypersurface that contains an invariant object. In particular, the hyperplane $\mathbf{X}(t) = 0$ is invariant. We assume that this rest state becomes unstable when the control parameter $p(\mathbf{Y}(t))$ exceeds a critical threshold $p_c$.

As $\mathbf{Y}(t)$ evolves, the control parameter $p(\mathbf{Y}(t))$ may fluctuate around the critical threshold $p_c$, occasionally entering regions where the rest state $\mathbf{X}(t) = 0$ is either stable or unstable. When these fluctuations are neither too brief nor too long, the system may display on–off intermittency: a pattern where $\mathbf{X}(t)$ remains near zero for irregular intervals and then suddenly exhibits transient bursts of large amplitude, driven by the instability induced during excursions above the threshold.

Another type of intermittency is known as  \textit{eyelet intermittency}. It can be detected at the boundary of the phase synchronization of coupled chaotic oscillators \cite{elaskar_new_2017}. Eyelet intermittency has been explained in terms of the synchronization of unstable periodic saddle orbits embedded in chaotic attractors of coupled oscillators \cite{kurovskaya_distribution_2008,rosa_transition_1998}.

Finally, \textit{crises-induced intermittency} occurs when the system undergoes one of the following two types of crisis \cite{grebogi_critical_1987}:
\begin{itemize}
    \item \textit{Attractor widening}. When the driving parameter exceeds a critical value, $p\ge p_c$, the single system attractor grows substantially, encompassing a larger and larger portion of the phase space. For $p\succsim p_{c}$, while the trajectory still spends a long time in the old region of the attractor (the one occupied by the original attractor before the widening), it is also subject to sudden jumps that cause it to enter the new available portion of the widened attractor, and then remains there for some time.
    
    \item \textit{Attractor merging}. Suppose that for $p<p_c$ there are two chaotic attractors, each with its own basin of attraction, which are separated by a basin boundary. As $p$ increases, the two attractors widen, and at the crisis ($p=p_c$), they simultaneously touch the basin boundary. For $p>p_{c}$, the orbit will alternate between the portions of the phase space originally occupied by the two separated attractors (now factually merged into a single one). The frequency of the alternation increases with $p>p_{c}$. 
\end{itemize}


\section{\label{sec:CLVs} Lyapunov exponents and vectors for characterizing intermittency}

The Lyapunov exponents (LEs) measure the average rates of expansion/contraction of perturbations of a trajectory, i.e. solution of a deterministic autonomous dynamical system like Eq.~\ref{eq:autonomo}. They are well-known quantities to characterize quantitatively and  qualitatively the behavior of a system, for example to determine the predictability horizon of the chaotic dynamics \cite{nayfeh_applied_1995}. 

The covariant Lyapunov vectors (CLVs) were first introduced by Ruelle in 1979 \cite{ruelle_ergodic_1979} under the name of characteristic Lyapunov vectors. The theory has then be further discussed and elaborated in the context of atmospheric dynamics by \citet{CLVsLegras}, and, \citet{Trevisan_1998}, but their efficient numerical computation has been made accessible only recently \cite{WolfeCLV,CLVComp,kuptsov_theory_2012,CLV,CLV2023}. They form a norm-independent basis for the tangent space that is covariant under the action of the linearized dynamics \cite{Vannitsem_2016}. Therefore, they can be computed once and then determined for all times using the Tangent Linear Model (TLM). The CLVs represent the individual direction of growth/decay of perturbations in the tangent space, whose rates are given by the corresponding local LEs. The CLVs can also be used to efficiently track and control the error in data assimilation \cite{Carrassi2022}. 

We shall use the LEs, the CLVs and other LE-based metrics to perform a global characterization of the intermittency. By global we mean here the study of the properties of the intermittency as the parameter varies, and therefore of the emergence of intermittency and its driving mechanisms. We will leverage and rely on the characterization of the intermittency detailed in Sec.~II. Our approach builds on, and it extends beyond, the bifurcation analysis and the statistical study of intermittency.

In parallel, we shall perform a local analysis of the intermittency. While, with the global analysis, we shall determine the conditions under which the system exhibits intermittency as a critical parameter varies; with the local view, our ultimate goal is to develop methods for identifying and predicting the regime transitions, at a fixed value of the parameter for which the system behaves intermittently. 

The computation of the CLVs and LEs that are presented in this work have been obtained by using the method from \citet{CLVComp}. The ones form Sec.~\ref{sec:Lorenz} to Sec.~\ref{sec:on-off} has been carried out with the support of the \textit{lyapunov} module within the \textsf{qgs} software \cite{Demaeyer2020}.

\subsection{Global analysis}

The spectrum of the LEs is used to associate qualitative asymptotic behavior of the system with the parameter values. In particular, we will identify portions of the parameter space where the systems display periodicity, chaotic behavior and the transition to it. From the LEs we shall compute the Kaplan-Yorke attractor dimension~\cite{FREDERICKSON1983185,GRASSBERGER1983189} ($d_{KY}$): the dimension of the invariant manifold in which the system is evolving or the fractal dimension of the strange attractor in the case of chaotic regimes. We shall also compute the Kolmogorov-Sinai entropy ($h_{KS}$) that measures the rate at which information about the initial conditions is lost over time. It is therefore used to characterize the degree of chaoticity of a system. Specifically, for $h_{KS}=0$ the system is not chaotic, while if $h_{KS}>0$ it exhibits chaotic behavior~\cite{chaosmesure}.

\subsection{Local analysis}

The CLVs form a covariant basis for the tangent linear space, which, at each point in the state space, can be decomposed into three subspaces: the unstable, neutral, and stable ones~\cite{Ginelli_2013,Vannitsem_2016}.
The former is represented by the CLVs associated to positive LEs and describes the directions of perturbations (errors) growth. The neutral subspace, typically aligned with the flow direction, contains CLVs with zero LEs. A well-known result established that all chaotic continuous dynamical systems of at least dimension three must possess one zero LE \cite{Pikovsky_Politi_2016}. We shall see in the following that some systems are degenerate in the neutral part of the spectrum, \textit{i.e.} they possess more than one zero LEs, and we will detail the implications of this degeneracy on the nature of their intermittent behavior. Neutral modes degeneracy has also been observed in prototypical atmosphere-ocean coupled dynamics \cite{tondeur_temporal_2020, Carrassi2022}.  
The stable subspace represents contracting directions.

This decomposition is typical of hyperbolic systems, which implies that the stable and unstable subspaces are never tangent to each other. Violations are ubiquitous, for example, through the occurrence of homoclinic tangencies, where the stable and unstable subspaces become tangent or near tangent to each other \cite{Ginelli_2013}.

The angles between the individual CLVs, or between some relevant subgroup of them, appear as a natural measure of the separation between the subspaces, thus offering valuable insights into the structure of the tangent space and the hyperbolicity of the system~\cite{Ginelli2007,Ginelli_2013,Vannitsem_2016,Critical2017}. This plays a key role on how perturbations grow, decay or remain unaltered. Such structural information can be relevant when the system undergoes transitions between different phases, which typically correspond to distinct regions of the state space characterized by different stability properties and structural features.

Computing angles among CLVs requires some caution. In fact, it is important to emphasize that while the CLVs are norm-independent and scalar-product-independent, the angles between them depend on the latter. In this work, we use the Euclidean scalar product. The cosine of the angle, $\theta_{ij}$, between the normalized CLVs $\mathbf{v}_i$ and $\mathbf{v}_j$ is computed according to 

\begin{equation}
    \label{eq:angle}
    \cos(\theta_{ij}) =<\mathbf{v}_i,\mathbf{v}_j>.
\end{equation}

Previous studies have discovered that CLV alignment along the flow direction offers a method to detect and predict high-amplitude peaks in certain chaotic dynamical systems \cite{CriticalTran, CLV_Alignment}. Other, more sophisticated, measures for CLVs alignment have been used in studying ensemble-based data assimilation \cite{Bocquet_Carrassi_2017}. 

In the recent work by~\citet{viennet_guidelines_2022}, the CLVs have been estimated directly from data and the approach tested in three different systems. Using a dynamical clustering method, they computed an approximation of the CLVs that is able to give information about the stability in different regions of the phase space. 
Similarly, in \citet{brugnago_machine_2020} the angles between the CLVs are used to train a neural network to predict the regime change in a geomagnetic dynamo model. 

The aforementioned previous studies share some aspects with our work, but differ in the following sense: Intermittent dynamics encompass a broad range of forms, which depend on the topological properties of the space in which the system evolves. For instance, in the case of type-I intermittency, the observed transition corresponds to a shift from a locally ordered, regular, state, to a chaotic one. This type of transition is structurally distinct from regime shifts occurring in, {\it e.g.}, metastable systems or in regions of the attractor associated with extreme events and high peaks. In this study, our objective is to investigate whether the alignment of CLVs can provide useful insights and information across a broader set of dynamical systems and intermittency mechanisms.

\section{\label{sec:Results} Results}

\subsection{\label{sec:Lorenz} Type-I intermittency in L63}

The well-known and celebrated L63 system \cite{lorenz_deterministic_1963} emerges from the study of Rayleigh-Bénard convection, the motion of a fluid layer heated from below and subject to gravity. The system describes the simplified behavior of the convective flow using three coupled nonlinear differential equations.

Following the pioneering work of Pomeau and Manneville \cite{pomeau_intermittent_1980}, the Type-I intermittency in L63 has then been widely studied \cite{elaskar_new_2017,MANNEVILLE19791,malasoma_multichannel_2003}. Here, the bifurcation parameter is $p=r$. The system exhibits a limit cycle for values of the Rayleigh number ranging from $r=148.4$ to $r=166.07$. When $166.07 <r\le 167$ the solutions alternate laminar periods, reminiscent of the limit cycle, with chaotic bursts on the other portion of the strange attractor. When $r>167$, the system transitioned to fully developed chaos. Thereafter, we numerically integrate the system with a RK4 numerical scheme and a time step of $\Delta t = 0.01$.

\subsubsection{Global view: Lyapunov exponents, covariant Lyapunov vectors and attractor dimension}

\begin{figure}[H]
    \includegraphics[width=0.45\textwidth]{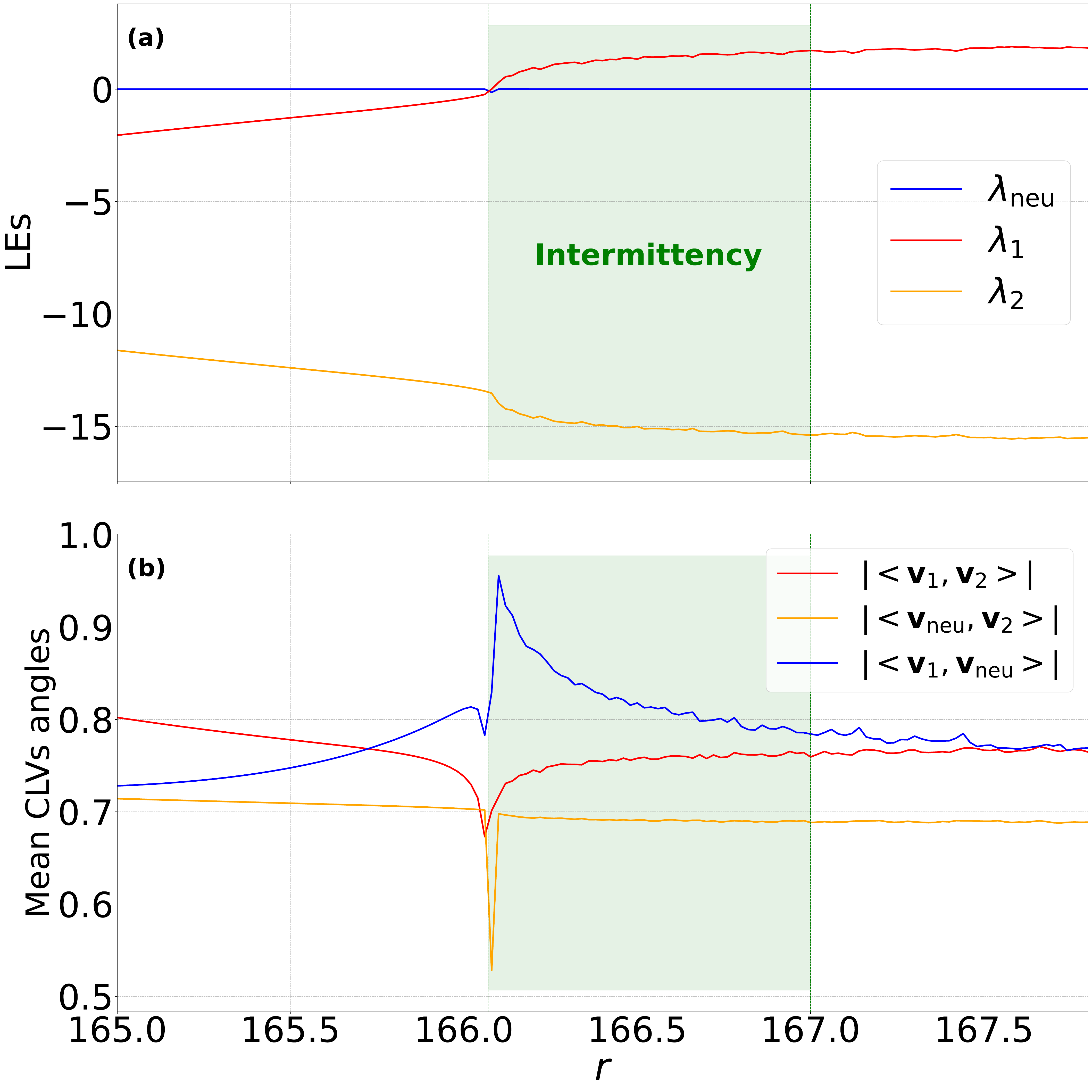}
    \caption{(a) Lyapunov exponents and (b) mean of the absolute value of the cosine of the angle between two CLVs. The green area represents the parameter range for which the system is intermittent.}
    \label{fig:lyap1}
\end{figure}

Figure~\ref{fig:lyap1} shows the LEs (Fig.~\ref{fig:lyap1}(a)) and the average cosine of the angle between pairs of CLVs (Fig.~\ref{fig:lyap1}(b)), as a function of the Rayleigh number at the regimes boundaries and within the intermittent regime.  Here, we use the notation $\lambda_{\rm neu}$ to indicate the exponent representing the neutral direction of the tangent space \footnote{It is the direction tangent to the flow.}, and $\lambda_1$ and $\lambda_2$ for the others. Similar notation is used for the CLVs. 

Before entering the intermittent region $(r<166.07)$, every LE $\lambda_i$ is non-positive. The largest LE is $\lambda_{\rm neu}=0$ consistent with the presence of a limit cycle.
For $(r\ge166.07)$, $\lambda_1$ becomes positive, indicating the transition to a chaotic regime, whereby $\lambda_1$ quantifies the average (asymptotic) rate of exponential growth of perturbations. 
$\lambda_{\rm neu}$ remains equal to zero for $r\ge166.07$, still indicating the neutral direction, tangent to the flow. The second LE slowly decreases with $r$ until $(r<166.07)$, counteracting the slow growth of $\lambda_1$. For $r\ge 166.07$, the decrease of $\lambda_2$ mirrors the growth of the leading LE, $\lambda_1$.

Figure~\ref{fig:lyap1}(b) displays the time average of the absolute value of the cosine of the angles between the three pairs of CLVs.  Notably, the three angles all show an abrupt change in correspondence to the beginning of the intermittency window. 

The Lyapunov exponents and the angles between the CLVs appear thus to both be able to detect the transition from the laminar to the intermittent regimes but are insensitive to entering the fully chaotic behavior. We argue that while the former transition is characterized by a strong and abrupt modification of the underlying phase space topology, a limit cycle turning into a strange attractor, the transition to fully developed chaos occurs through smoother changes of the attractor and of the tangent space splitting. 

\begin{figure}[H]
    \includegraphics[width=0.42\textwidth]{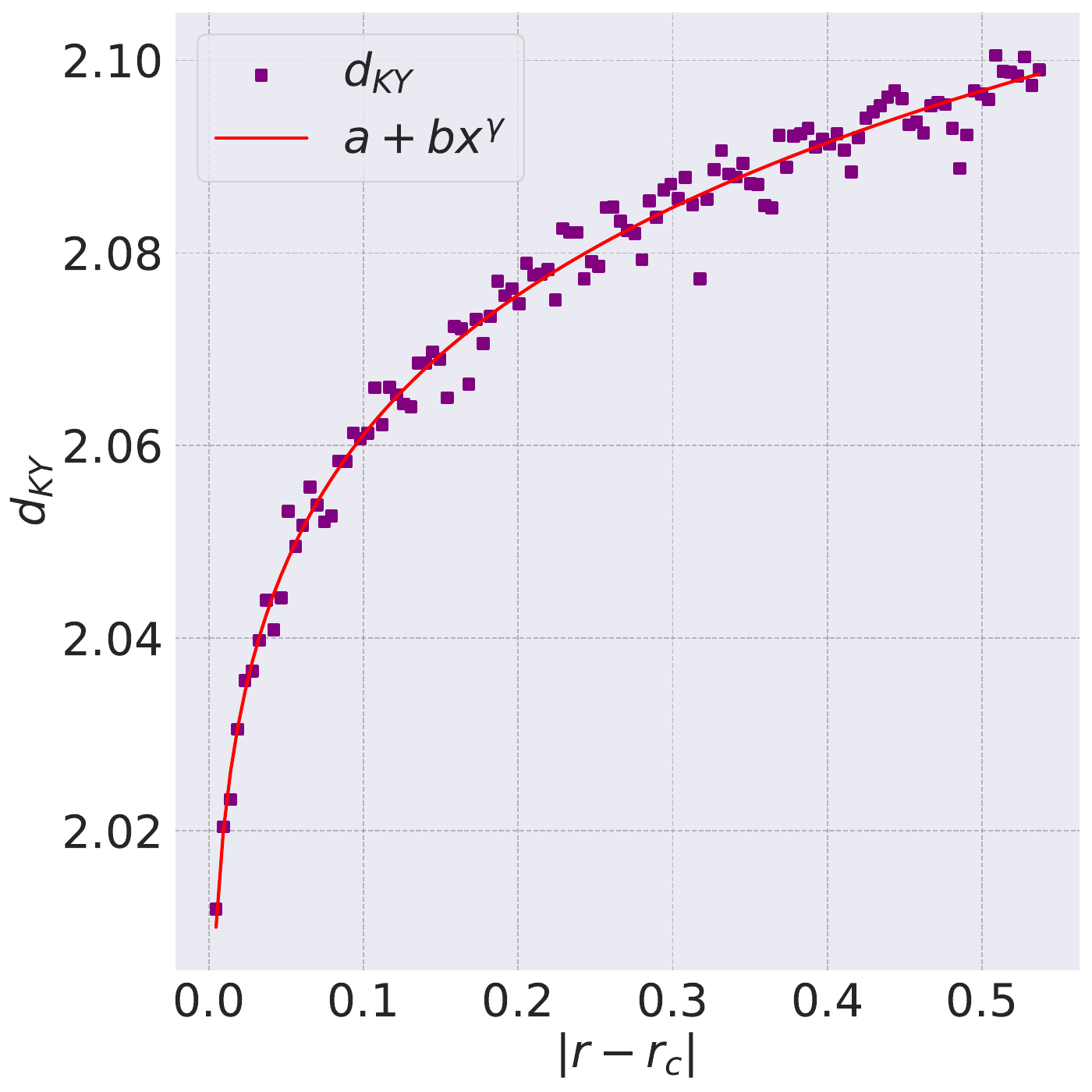}
    \caption{Kaplan-Yorke dimension as function of the deviation from the critical value ($r_c=166.06$) of the bifurcation parameter. In red is shown the best, power-law like, fit to the data.}
    \label{fig:KYdim}
\end{figure}

As highlighted by Pomeau and Manneville \cite{MANNEVILLE19791}, the transition from a periodic to an intermittent regime in type-I intermittency can be understood as a shift from a limit cycle to a strange attractor. The hypothesis is that the expansion of the attractor is driven by an increasing number of chaotic bursts. The strange attractor then gradually expands as the critical parameter increases, while still preserving the limit cycle as an underlying substructure. To corroborate this conjecture, we show in Fig.~\ref{fig:KYdim} the fractal dimension of the attractor, the Kaplan-Yorke dimension ($d_{KY}$), as a function of the deviation from the critical value of the parameter.

We see that the fractal dimension of the attractor increases within the intermittent window, following a power law:
\begin{equation}
    d_{KY} \propto |r-r_c|^\gamma,
    \label{eq:dfit1}
\end{equation}
with $\gamma \thickapprox 0.1 $.

In the neighborhood of the inverse tangent bifurcation of the local map, when $\varepsilon = |r-r_c|  \gtrsim 0$, we can use Eq.~\ref{eq:I8}:
\begin{equation}
    \overline{l} \propto |r-r_c|^{-\frac{1}{2}}.
    \label{eq:avgll}
\end{equation}

From Eqs.~\eqref{eq:dfit1} and \eqref{eq:avgll}  we obtain an expression linking the fractal dimension of the attractor to the average laminar length 
\begin{equation}
    d_{KY} \propto \frac{1}{\overline{l}^{2 \gamma}}.
     \label{eq:dimll}
\end{equation}

Equation~\ref{eq:dimll} is extremely powerful. It states that for type-I intermittency, the fractal dimension of the attractor scale with average laminar length. The latter can be estimated numerically, albeit subject to potentially high computational costs.

\subsubsection{Local view: do the CLVs indicate the regimes' change?}

\begin{figure*}
    \centering
    \includegraphics[width=0.8\textwidth]{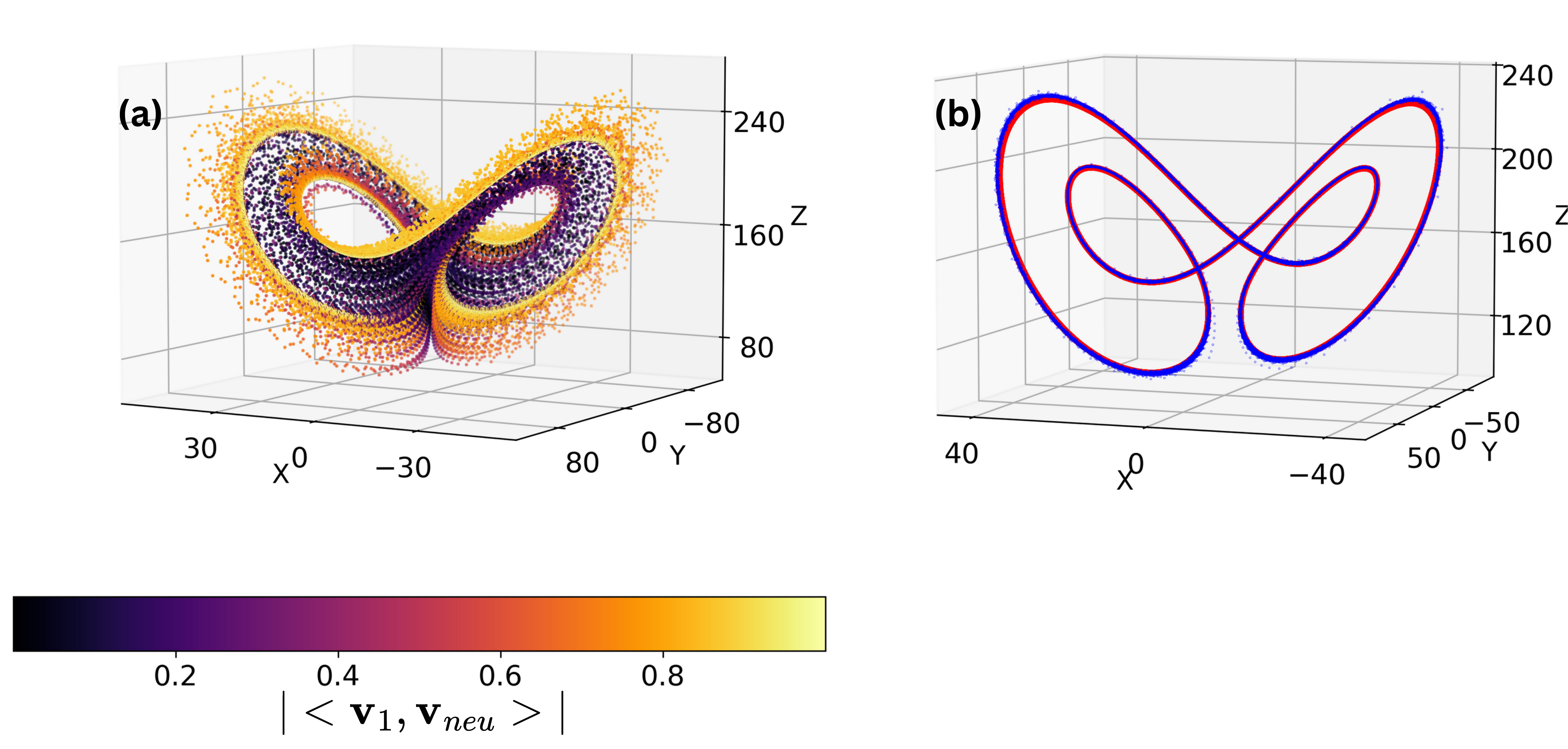}
    \caption{(a) Absolute value of the moving mean (right aligned and 50 time steps lengths) of the cosine between $\mathbf{v}_{1}$ and $\mathbf{v}_{\rm neu}$. (b) Limit cycle for $r=166$ (in red) and points sampled from the attractor when the two vectors are perfectly collinear (in blue.}
    \label{fig:angles}
\end{figure*}

We now fix the controlling parameter to $r=166.2$, so that the system behaves intermittently, and investigate the capabilities of CLVs to indicate or possibly predict regime changes. In Fig.~\ref{fig:angles}(a), we show the absolute value of the moving mean (right-aligned and 50 time steps length) of the cosine of the angle between the first CLV, $\mathbf{v}_{1}$,  and the neutral one, $\mathbf{v}_{\rm neu}$. The values are displayed on top of the system's trajectory, and reveal clearly two district stripes in the attractor, one characterized by a strong alignment of the two CLVs (cosine near $1$) and another by quasi orthogonality. Notably, the boundaries between the two stripes reflect the structure of the limit cycle present before the intermittent window.  

\begin{figure}[H]
    \includegraphics[width=0.48\textwidth]{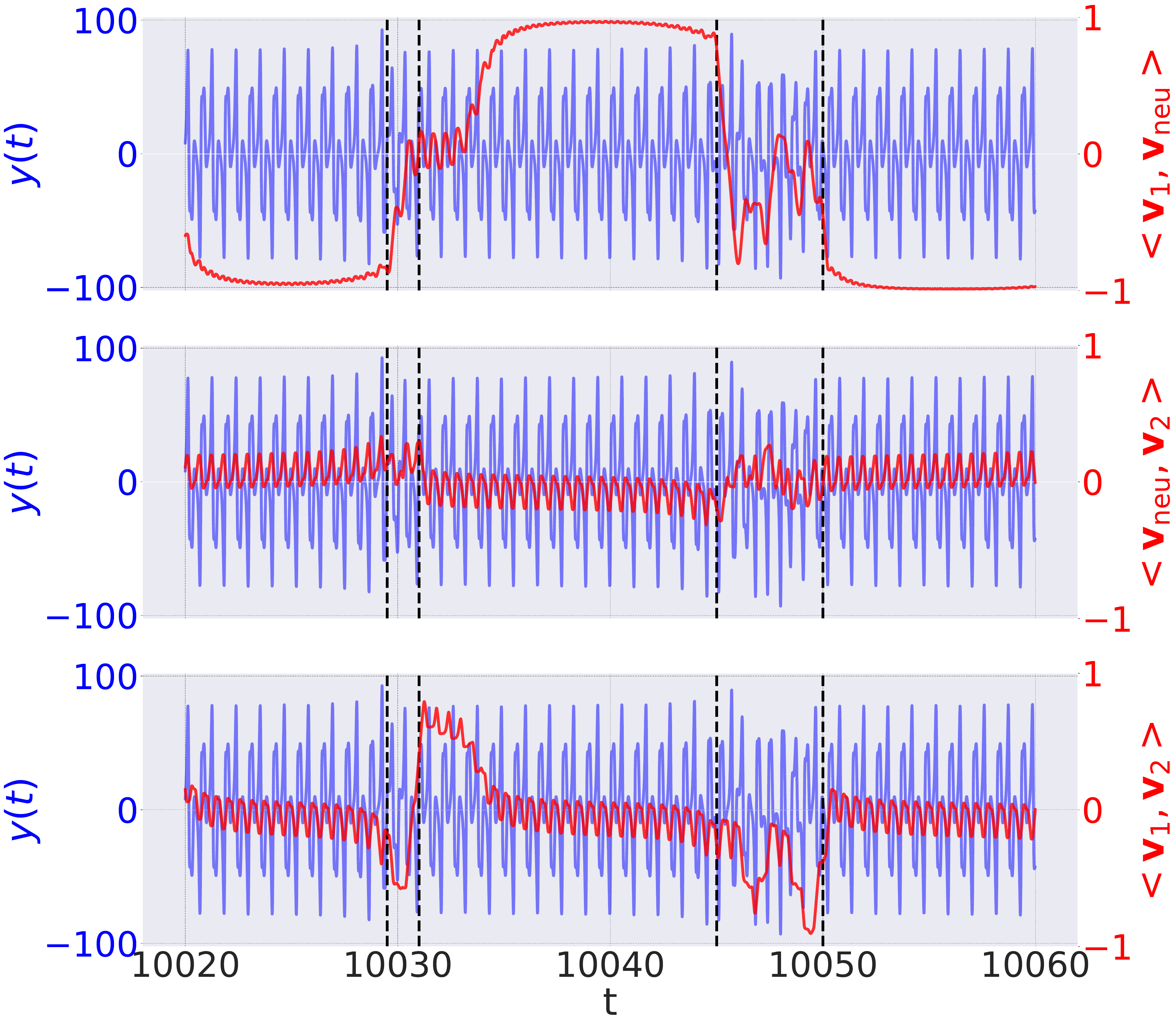}
    \caption{Intermittent time series of the $y$ component of the state vector for $r=166.2$ (in blue) and right aligned running mean of the angles between the CLVs (in red). The horizontal black dashed lines show the beginning and the end of the chaotic burst.}
    \label{fig:rollTS}
\end{figure}

In Fig.~\ref{fig:rollTS}, we show smoothed time series with a right-aligned moving mean of $100$ time steps for the three angle cosines between pairs of CLVs (red lines), and the time series of the $y$ variable (blue lines). The time interval considered (10020--10060) has been chosen arbitrarily and is representative of any equally long periods. 
It is noteworthy how $\mathbf{v}_1$ and $\mathbf{v}_{\rm neu}$ lose collinearity ($<\mathbf{v}_1,\mathbf{v}_{neu}>\approx0$) during the chaotic bursts (contained by two successive dashed vertical lines). 

\begin{figure}[H]
    \centering
    \includegraphics[width=0.4\textwidth]{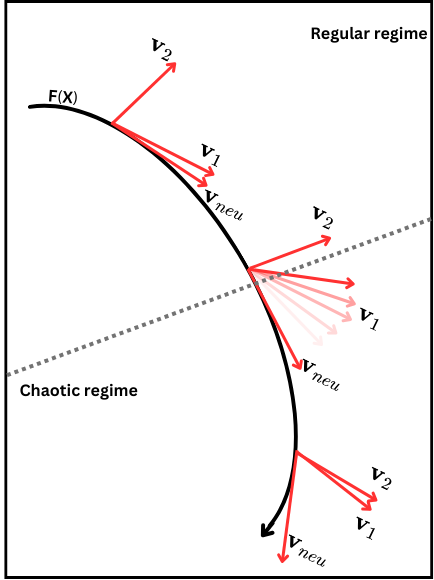}
    \caption{Illustration of the CLVs configuration, and how they change through, in the laminar and chaotic phase.}
    \label{fig:schemevec}
\end{figure}

The angle cosine between $\mathbf{v}_{{\rm neu}}$ and $\mathbf{v}_2$ is, on average, close to zero (cf Fig.~\ref{fig:lyap1} and \ref{fig:rollTS}): they are almost orthogonal to each other. This orientation oscillates regularly in the laminar phase but becomes irregular during the chaotic burst. The angle cosine between $\mathbf{v}_1$ and $\mathbf{v}_2$ behaves symmetrically to $<\mathbf{v}_1,\mathbf{v}_{\rm neu}>$: during the chaotic burst, the vectors are more collinear, while in the laminar phase, they oscillate around orthogonality. Overall, the first vector aligns with the neutral direction in the laminar phase and with the stable direction during the chaotic burst, while the neutral and stable directions remain less collinear, no matter the regime. An illustration of this mechanism is given in Fig.~\ref{fig:schemevec}. 

The evident connection between the relative directions of the CLVs and the type of dynamical regimes (laminar or chaotic) motivates the application of a classification algorithm. To this end, let first transform the time series of 
$<\mathbf{v}_1,\mathbf{v}_{\rm neu}>$ and $y(t)$  into square signals, with the values of one and zero corresponding to the laminar and chaotic phase respectively. We then treat the state variable signal as the observation and the signal from the angle as the forecast. We construct a contingency table \cite{wilks_forecast_2011} and count the outputs. The ``hit'' (H) is when both observation (the state TS) and forecast (CLVs signal) is present, {\it i.e.} when both indicate a burst; the ``miss'' (M) when there is the burst but it is not present in the forecast, the ``false alarm'' (FA) when a burst is predicted but it is not present in the observation and finally the ``correct rejection'' (CR) when the burst is absent in both signals. The total number of events is given by $n = H + M + FA + CR$. One can furthermore define the total success, also called ``proportion correct'' $PC = (H+CR)/n$ and the ``total fail'' $F = (M+FA)/n$. We obtain a total success of $PC= 90.76$, with relative $H=43.11$ and $CR=56.89$; and a total fail of $F=9.24$, with relative $M=33.66$ and $FA=66.34$. We also compute the \textit{Yule's Q} \cite{YuleQ} skill score $Q = (H \cdot CR -M\cdot FA)/(H \cdot CR +M\cdot FA) = 0.98$, showing that the angle performs very well in indicating the two different states.
To further confirm this, we compute the distribution of the laminar lengths independently from the two square signals. The two distributions (not shown) are very similar with a Wasserstein distance between them as small as $0.54$.

As mentioned before, the limit cycle remains as a substructure of the attractor, visible as the boundary between the two stripes highlighted by the cosine between the CLVs (see Fig.~\ref{fig:angles}).
Can we reconstruct the limit cycle (its reminiscence thereof) by using the information from only one of these angles?

To answer this, we consider the angle between the first and neutral CLVs, $<\mathbf{v}_1,\mathbf{v}_{\it neu}>$; similar results (not shown) are obtained by considering instead $<\mathbf{v}_1,\mathbf{v}_2>$. In contrast, using $<\mathbf{v}_2,\mathbf{v}_3>$ would have been more complex due to the reduced clarity in the separation of the two stripes. 
In Fig.~\ref{fig:angles}(b) we draw the pre-intermittent limit cycle for $r=166$
(red curve). We then superimpose points when $<\mathbf{v}_1,\mathbf{v}_{\rm neu}> \approx 1$ (blue). The similarity is spectacular: the blue points concentrate in the region of the limit cycle, with only few outliers corresponding to the oscillations observed during the bursts, as shown in Fig.~\ref{fig:rollTS}. 
We compute the Wasserstein distance for each normalized components of the two structures: $WD(x) = 0.02$, $WD(y) = 0.02$, $WD(z) = 0.01$. To assess whether the distributions of each component differ significantly between the two structures, we perform a two-sample Kolmogorov–Smirnov test. In all three cases, the resulting p-values are greater than $0.05$, indicating that there is no statistically significant difference between the corresponding distributions. This suggests that despite the presence of some outliers during burst oscillations, the overall statistical properties of the components remain consistent with those of the limit cycle.

Overall, for Type-I intermittency in L63, the CLVs have proven to be an excellent indicator for the transitions from laminarity to chaos. They allow for almost perfectly reconstructing the underlying limit cycle existing before the intermittency window. 

\subsection{\label{sec:Greg} Merging-crisis intermittency - The double well potential system}

A prototypical minimal model displaying merging-crisis intermittency is given by

\begin{equation}
  \frac{\mathrm{d}^2x}{\mathrm{d}t^2} + \nu \frac{\mathrm{d}x}{\mathrm{d}t} + \frac{\mathrm{d} V}{\mathrm{d} x} = F(t),
    \label{eq:oscillator}
\end{equation}

which represents an oscillator, with $V(x)$ being the potential energy and $F(t)$ a periodic forcing. 
The chaotic behavior of this class of systems has been widely explored by \citet{noauthor_nonlinear_1979,arecchi_low-frequency_1984, steeb_chaotic_1986}. We skip the global analysis and choose $F(t)=p \sin(\omega t)$ and $V(x)= \alpha x^4/4 - \beta x^2/2$, with $\alpha=100$, $\beta=10$, $p=0.855$ and $\omega = 3.5$. With those choices, it has been shown that the system exhibits merging crises intermittency \cite{symchaos, grebogi_critical_1987}. Eq.~\eqref{eq:oscillator} is a 2nd-order, non-autonomous ODE, which can be conveniently rewritten as a system of 1st-order autonomous ODEs defined in the cylindrical manifold $\mathbb{S}^1 \times \mathbb{R}^2$ 

\begin{equation}
\begin{cases}
  \dot{X} = -\nu X - \alpha Y^3 + \beta Y + p \sin (\omega Z) \\
  \dot{Y} = X \\
  \dot{Z} = 1.
\end{cases}
    \label{eq:gregsys}
\end{equation}

\begin{figure}[H]
    \centering
    \includegraphics[width=0.5\textwidth]{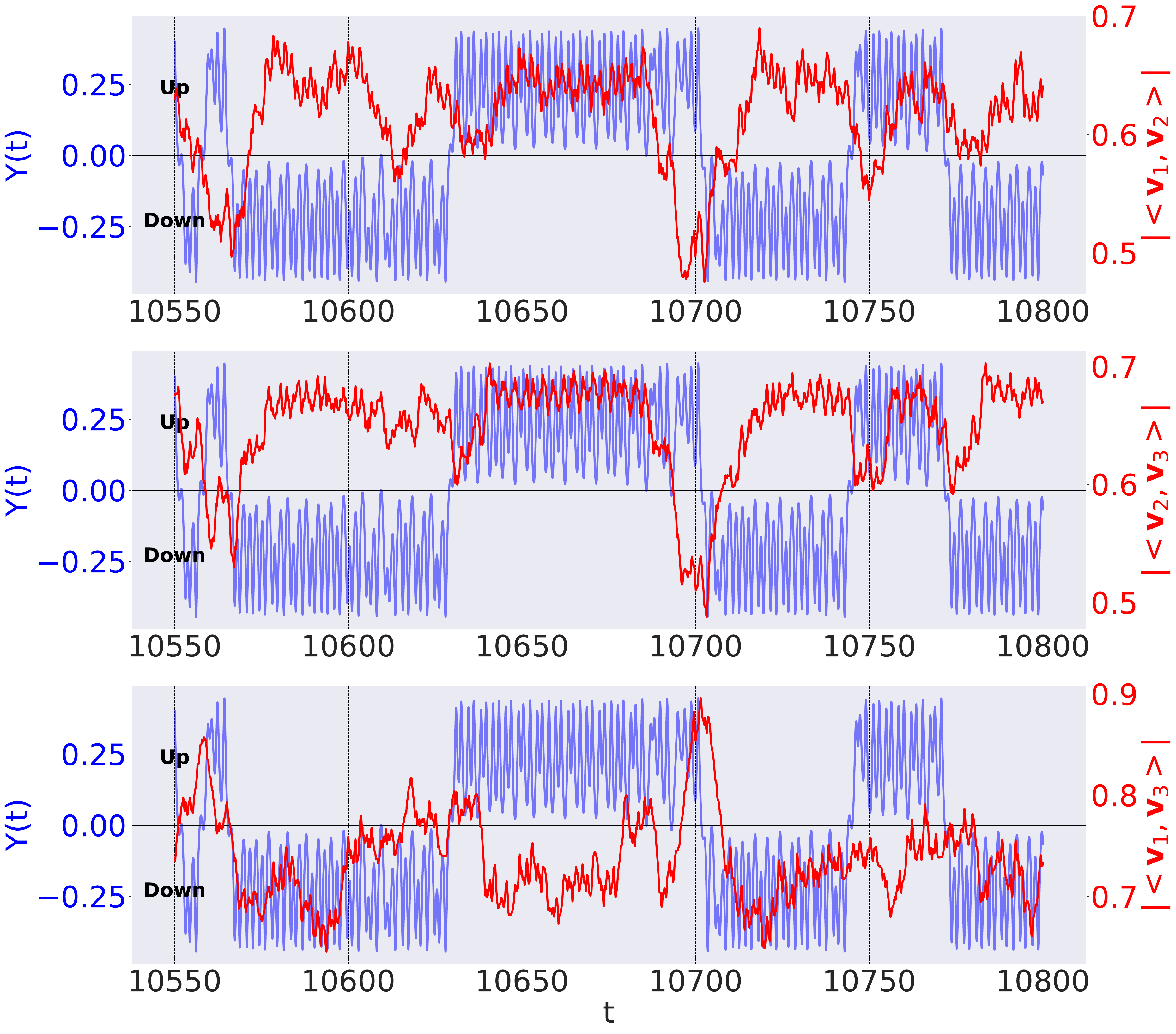}
    \caption{Intermittent time series of the component $Y$ of the state vector (blue) and of the angles between pairs of CLVs (red).}
    \label{fig:CLVgreg}
\end{figure}

We integrate Eq.~\eqref{eq:gregsys} using RK4 and a time step of $\Delta t = 0.01$, and compute the CLVs and the absolute value of the cosine of the relative angles between pairs of them. We then apply a moving average to $|<\mathbf{v}_i,\mathbf{v}_j>|$, with a right-aligned window of $1,000$ time steps. Here $\mathbf{v}_1, \mathbf{v}_2$ and $\mathbf{v}_3$ are respectively the unstable, neutral and stable CLV.  The results are shown in Fig.~\ref{fig:CLVgreg}.

We can see that $|<\mathbf{v}_1,\mathbf{v}_2>|$ and $|<\mathbf{v}_2,\mathbf{v}_3>|$ are capable of indicating the transition from the positive (``up'' state) to the negative (``down'' state) of the state variable $Y(t)$. In particular, we observe that the vectors remain closer to each other when the system is either up or down, and lose collinearity only at the moment of the transition.

\subsection{\label{sec:L96} Merging crises intermittency with periodic intermission - L96 system}

\begin{figure*}
    \includegraphics[width=\textwidth]{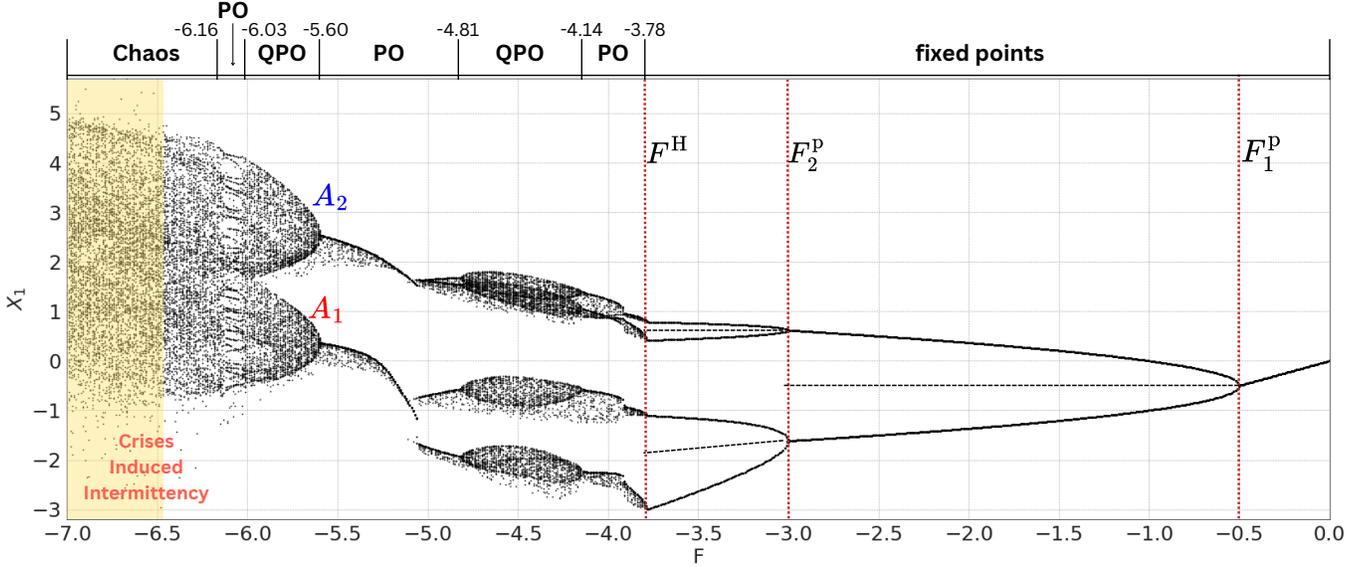}
    \caption{Bifurcation diagram for the L96 system, including additional information from existing literature and new numerical observations.}
    \label{fig:sumbif}
\end{figure*}

The Lorenz 96 (L96) model \cite{Lorenz_2006} is described by the following set of equations:

\begin{equation}
\dot{X}_j = (X_{j+1} - X_{j-2}) \cdot X_{j-1} - X_j + F, \quad j = 1, 2, \ldots, N,
\label{lorenz96}
\end{equation}

with boundary conditions:

\begin{equation}
    X_{j-n} = X_{j+n} = X_j .
\end{equation}

The state variables have historically been interpreted as a generic meteorological quantity, along one loop at a fixed latitude, so that the $j = 1, 2, \ldots, N$ indices represent the longitude. Despite its simplicity, this model is able to capture the essential structure of an atmospheric model. Its dimensionality can also be easily modified by adding new variables $X_j$, allowing one to obtain arbitrarily large configurations. All these features have made the L96 an extremely popular model in many applications, from data assimilation and machine learning \cite{TrevL96, MLL96} to statistical mechanics \cite{Luc_2011}.

\subsubsection{Global analysis: bifurcations and intermittency regions}

As in the case of L63, we start with a global approach, examining how the dynamical properties of the system change as the forcing $F$ varies. Building on the work of \citet{VKSTERKL962019} (2019), our analysis focuses on the negative part of the parameter interval. \citet{VKSTERKL962019} gave a general view of the behavior of the system for $F<0$, by using the equivariant bifurcation theory. In particular, they showed that for any $n \in \mathbb{N}$, the L96 system is equivariant under cyclic left shift of the coordinates, a property known in bifurcation theory as a $\mathbb{Z}_n$ symmetry, since the left shift cyclic group is isomorphic to $\mathbb{Z}/n\mathbb{Z}$. 
This equivariant property implies the emergence of invariant linear subspaces and \citet{VKSTERKL962019} showed how these subspaces can be used to determine some specific behaviors of the systems. They classified three different structures for the system for $n \ge 4$ and $F<0$:

\begin{itemize}
    \item $n$ odd, the first bifurcation for the equilibrium point $x_*= (F, F, ... , F)$ is a supercritical Hopf bifurcation
    \item $n=4 k + 2$, with $k \in \mathbb{N}$, the first bifurcation is a pitchfork bifurcation and the systems undergo a Hopf bifurcation on each branch. 
    \item $n=4 k$, with $k \in \mathbb{N}$, a first pitchfork bifurcation (PB) occurs at $F^{\rm{p}}_{1} = -1/2$, and then each branch shows a pitchfork bifurcation at $F^{\rm{p}}_2 = -3$. Finally all four stable equilibrium of the second PB exhibit a Hopf bifurcation, which happens between $-3.9<F^{\rm{H}}<-3.5$.
    
\end{itemize}

Our investigation focuses on the search for intermittency windows for a configuration of the system that falls within the third case ($n=8, F<0$). We integrate the system with a RK4 integration scheme and a time step of $\Delta t = 0.01$. In order to find intermittency windows, we make a bifurcation diagram that builds on and extends the work by \citet{VKSTERKL962019}; results are given in Fig.~\ref{fig:sumbif} which put together the original results by \citet{VKSTERKL962019} and our extension. Our numerical results are in agreement with \citet{VKSTERKL962019}, but furthermore allow us to identify a part of the bifurcation diagram in which two attractors($A_1$ and $A_2$) coexist and individually present a periodic window for $-6.16<F<-6.03$. By further reducing the forcing (increase in absolute value), we can then notice that the two attractors widen and start to encompass each other, {\it i.e.} an attractor merging crisis is taking place. This gives rise to a crises-induced intermittency \cite{grebogi_critical_1987}. We thus face a very peculiar situation in which each attractor individually is passing from a periodic window to a chaotic one while simultaneously experiencing a merging crisis. 

The Kolmogorov-Sinai entropy ($h_{KS}$) and the Kaplan-Yorke ($d_{KY}$) dimension have been computed for one of the possible coexisting attractors (not shown): $h_{KS}$ indicates that the system becomes chaotic on the left of the periodic window for $F \approx -6.16$; $d_{KY}$ tells us that until the second four simultaneous Hopf bifurcation (for $F^{\rm{H}}$), the dimension is $d_{KY}=0$. This is clearly due to the fact that, after the transient, the system stabilizes into one of the stable fixed points of the pitchfork bifurcation. After the Hopf bifurcation at $F^{\rm{H}} \approx -3.78$ we observe the four simultaneous periodic orbits (POs) \cite{VKSTERKL962019}. 
This PO interval lasts until the space explored by the system, for $-4.81<F<-4.14$, has dimension two. The system densely covers four two-dimensional invariant torus. 

At $F=-4.81$ the system passes again to four simultaneous periodic orbits that become two at $F=-5.06$.
In the interval $-6.03<F<-5.60$ the system explores again a two-dimensional space,  two simultaneous two-dimensional invariant torus, which are the $A_1$ and $A_2$ mentioned before. 
Then we find the periodic window ($-6.16<F<-6.03$) mentioned above, and for $F<-6.16$ the system transitions from the POs to the two simultaneous attractors.
Finally, for $F \approx -6.4717$ we identify an attractor merging, causing the crises-induced intermittency. 

We compute the scaling of the crises-induced intermittency against the deviation from the critical value of the forcing. Specifically, we compute the average time the system spends in each region of the state space where the two attractors were present before the crises $F>F_{mc}$. We then fit the data using the well-known scaling law $\tau \propto |F-F_{c}|^{\gamma}$, obtaining an exponent of $\gamma \approx -2.17$ (see Supplementary Material).

\subsubsection{Local analysis}

\begin{figure}[H]
    \centering
    \includegraphics[width=0.48\textwidth]{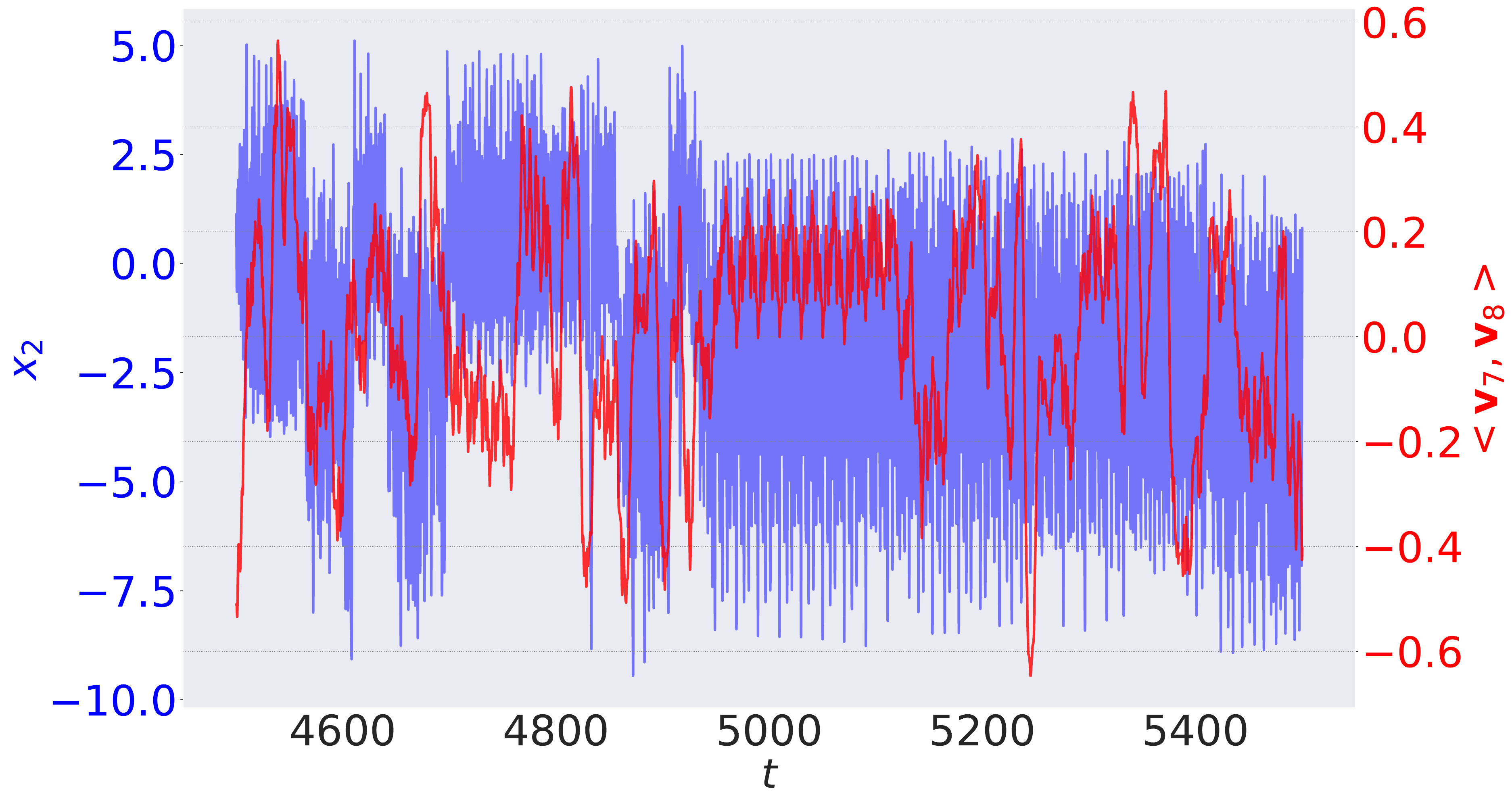}
    \caption{Time series for the state variable $x_2$ (blue) and smoothed time series of $<\mathbf{v}_7, \mathbf{v}_8>$ (red).}
    \label{fig:L96TS}
\end{figure}

Having identified the interval in which we have crisis-induced intermittency, we perform the local analysis, fixing the forcing at $F=-7$. In Fig.~\ref{fig:L96TS} we can observe in blue the time series of $x_2$ (the same qualitative behavior is found in other components) characterized by the typical jumps due to the ``crisis''. As per previous figures, we superimpose the smoothed time series of $<\mathbf{v}_7,\mathbf{v}_8>$; a moving average with a right-aligned window of 1000 time steps has been used. This pair of vectors was chosen because it provided the best performance in this specific case. We can also see a periodic intermission starting around $t=4,950$, and ending just before $t=5,100$. Periodic intermissions of this kind are due to the aforementioned existence of periodic windows in each of the attractors $A_1$ and $A_2$, before reaching the crisis. 

In this system configuration, the CLVs effectively unveil the presence of the periodic intermissions, as they begin to oscillate in a regular manner during these phases. This regularity reflects the temporary re-emergence of the underlying periodic dynamics, making the intermissions clearly distinguishable from the surrounding chaotic behaviors. This suggests that the predominant dynamical feature of the system is the switching between the periodic and chaotic phases rather than the jumps induced by the merging crises. Such a conjecture is supported by the fact that the CLVs successfully capture the regime shifts in the case of the double-well potential system and in other configurations of the L96 systems where the periodic intermissions are not present (not shown here). 

We performed extensive experiments with different configurations of L96 with negative values of the forcings. In all cases, whenever $n = 4k+2$ or $n = 4k$ $k\in\mathbf{N}$, {\it i.e.} those featuring at least one PF bifurcation, it leads to a merging crises. We hypothesize that this pattern can be generalized beyond those we investigated numerically, to all system configurations with $n = 4k+2$ and $n = 4k$. We believe that this is due to the emergence of the first pitchfork bifurcation, that leads to the formation of the two branches that will eventually evolve into chaotic attractors.

\subsection{\label{sec:on-off} On-off intermittency - Platt-Spiegel system}

\begin{figure*}
    \includegraphics[width=\textwidth]{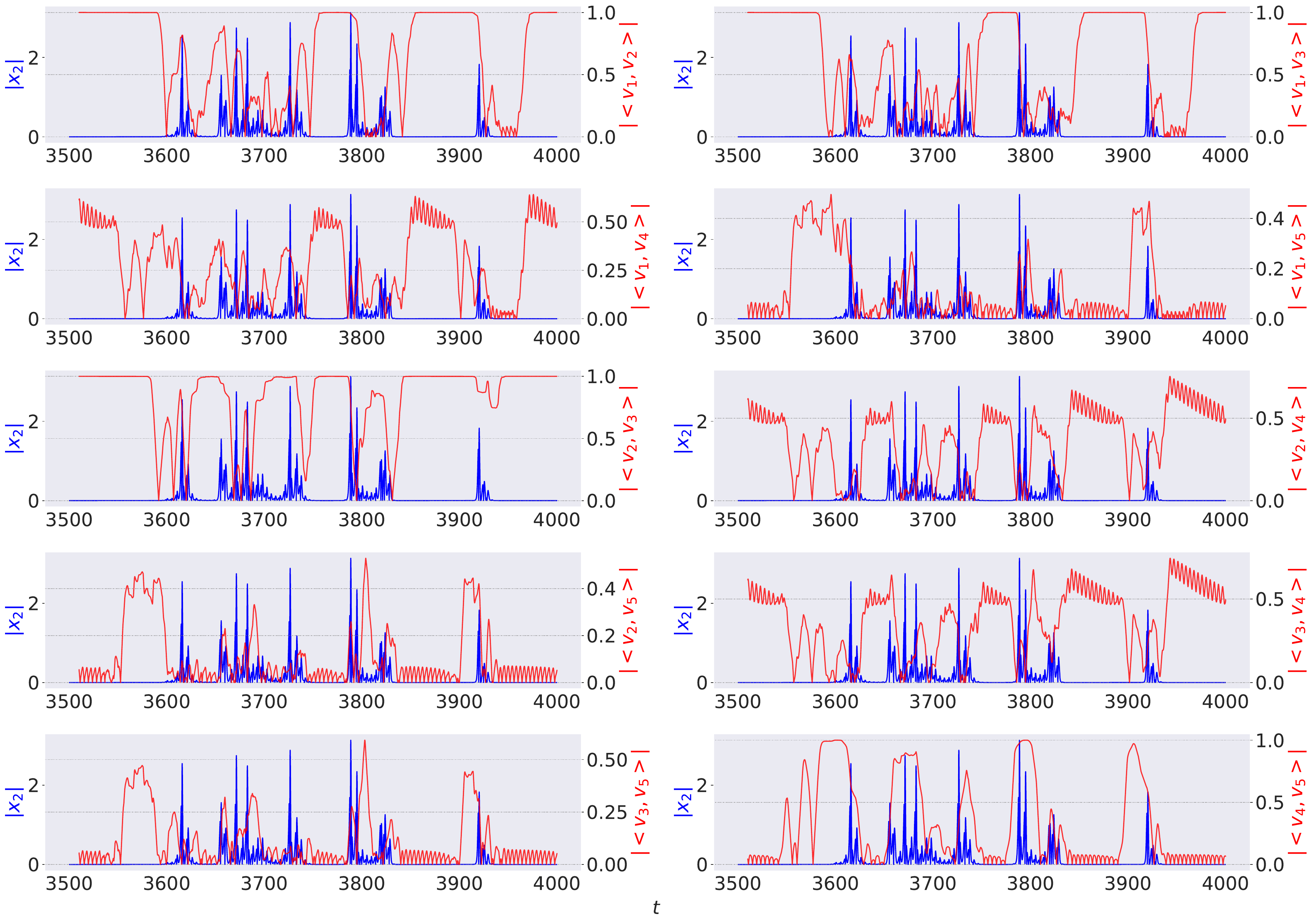}
    \caption{Absolute value of the state variable $x_2$ (blue) and absolute value of all the combinations $<\mathbf{v}_i,\mathbf{v}_j>$ with $i,j=1,...,5$ and $i \neq j$ (red).}
    \label{fig:clvsonoff}
\end{figure*}

We now consider a simple system for on-off intermittency~\cite{platt_-off_1993}: 

\begin{equation}
\begin{cases}
    \dot{x_1} = x_2 \\
    \dot{x_2} = -x_1^3 -2x_1 x_3 + x_1 x_5 - \mu_{01} x_2 \\
    \dot{x_3} = x_4 \\ 
    \dot{x_4} = - x_3^3 - \nu_{01} x_1^2 + x_5 x_3 - \nu_{02} x_4  \\
    \dot{x_5} = -\nu_{03} x_5 - \nu_{04} x_1^2 - \nu_{05} (x_3^2 -1) .\\
\end{cases}
\label{eq:platt_sys}
\end{equation}

The system possesses a similar structure to Eq.~\ref{eq:platt_sysa} with $N=2$, $S=5$ with the hyperplane $X=0$ being an invariant. We integrate the system with a RK4 integration scheme and a time step $\Delta t=0.01$. The parameters are $\mu_{01}=1.815$, $\nu_{01}=1$, $\nu_{02}=1.815$, $\nu_{03}=0.44$, $\nu_{04}=2.86$, $\nu_{05}=2.86$, for which the system shows on-off intermittency. Note that, by setting $x_1 = x_2 = 0$, we recover a third order system that is a version of L63 \cite{lorenz_deterministic_1963}. With this configuration the system is chaotic with $\lambda_1 = 0.111$.

In Fig.~\ref{fig:clvsonoff} we show the smoothed time series of the cosine of the angles among pairs of CLVs. The smoothing has been performed using a moving average with a right-aligned window of $1,000$ time steps. It is evident that almost all combinations provide insight into regime changes. 
Notably, the combination that performs most effectively is the angle between the last two CLVs. This angle not only exhibits a peak during the ``on'' phase of the dynamics but also oscillates near zero during the ``off'' phase. 
It is also noteworthy that the end of this oscillation systematically anticipates the burst to the ``on'' phase. This suggests that the signal could potentially serve as a precursor of regime changes. Interestingly, whenever the most stable direction ($\mathbf{v}_5$) is considered, the oscillatory behavior is consistently observed during the ``off'' phase. 

Assuming that these ``off'' phases observed in the on-off intermittency are representative of a ``laminar'' state of the system, this observation suggests a potentially general mechanism pointing to a strong connection between stable directions and laminar phases. This is further supported by the findings in Sec.~\ref{sec:L96}, where a similar behavior was identified during the laminar intermissions.

\subsection{Spatially extended systems - The Kuramoto-Sivashinksy equation}

The Kuramoto-Sivashinsky equation (KSE), 
\begin{equation}
    u_t+u_{xx}+ \nu u_{xxxx}+uu_x = 0 ,
   \label{eq:KS1}
\end{equation}

with periodic boundary condition: 

\begin{equation}
    u(x,t) = u(x+L,t) ,
   \label{eq:BC}
\end{equation}
has been derived to describe the amplitude of interfacial instabilities in various physical conditions \cite{KSE1}. Here, we use the notation $u_x$ to indicate the partial derivative of $u$ with respect to $x$. Kuramoto and Tsuzuki \cite{Kuramoto1976PersistentPO} introduced it in the study of angular-phase turbulence within reaction-diffusion systems. Similarly, Sivashinsky \cite{SIVASHINSKY19771177} derived it to describe small thermal-diffusive instabilities in laminar flame fronts in two dimensions. 
The KSE is among the simplest nonlinear PDE displaying spatio-temporal chaos \cite{Steph_Nicolis1994,manneville2005dissipative}. 

In the KSE, both the spatial length $L$ and the viscosity $\nu \in \mathbb{R}^+$ play the role of bifurcation parameter. We fixed $L=10 \cdot \pi$ so that $x \in [0,L]$, and study the system's behavior for varying $p=\nu$. We search for windows in the viscosity range for which KSE is intermittent.

But first the periodic nature of the spatial domain prompts us to work in a more convenient framework, and to rewrite the KSE in terms of Fourier modes \cite{KSCvi}: 

\begin{equation}
\begin{split}
    \hat{u}(x,t) &= \mathcal{F}[u]_k = \frac{1}{L} \int_0^L u(x,t) e^{-iq_kx} dx, \\
    u(x,t) &=  \mathcal{F}^{-1}[\hat{u}] = \sum_{k \in \mathbb{Z}} \hat{u}_k e^{-iq_kx} .
   \label{eq:BC}
\end{split}
\end{equation}

The KSE becomes: 

\begin{equation}
    \frac{{\rm d}\hat{u}_{k}}{{\rm dt}} = \left(q_{k}^{2} - q_{k}^{4}\right)\hat{u}_{k} - \frac{iq_{k}}{2}\mathcal{F}_{N}\left[\left(\mathcal{F}_{N}^{-1}\left[\hat{u}\right]\right)^{2}\right]_{k}  ,
\end{equation}

with the associated equation in the tangent space given by:

\begin{equation}
    \frac{{\rm d}\hat{w}_{k}}{{\rm dt}} = \left(q_{k}^{2} - q_{k}^{4}\right)\hat{w}_{k} - iq_{k}\mathcal{F}_{N}\left[\mathcal{F}_{N}^{-1}\left[\hat{u}\right] \otimes \mathcal{F}_{N}^{-1}\left[\hat{w}\right]\right]_{k}. 
\end{equation}

These stiff ODEs are solved using the fourth-order exponential time differencing method ETDRK4 as proposed by Cox and Matthew \cite{COX2002430, kassam_fourth-order_2005, KSBer} with a time step of $\Delta t = 0.1$. The spatial domain is discretized in $n=101$ nodes.

\subsubsection{Global analysis \label{sec:KSEglobal}}

We start by analyzing the qualitative behavior of the system as the viscosity $\nu$ varies. We observe that the system exhibits chaotic behavior for $\nu \in (0, 0.855)$. An intermittent window is identified for $\nu \in [0.855, 0.865)$, which subsequently leads to periodic behavior for $\nu \in [0.865, 3)$. Furthermore, we highlight that the system stabilizes at a fixed point for $\nu > 29$.

\begin{figure*}
    \centering \includegraphics[width=\textwidth]{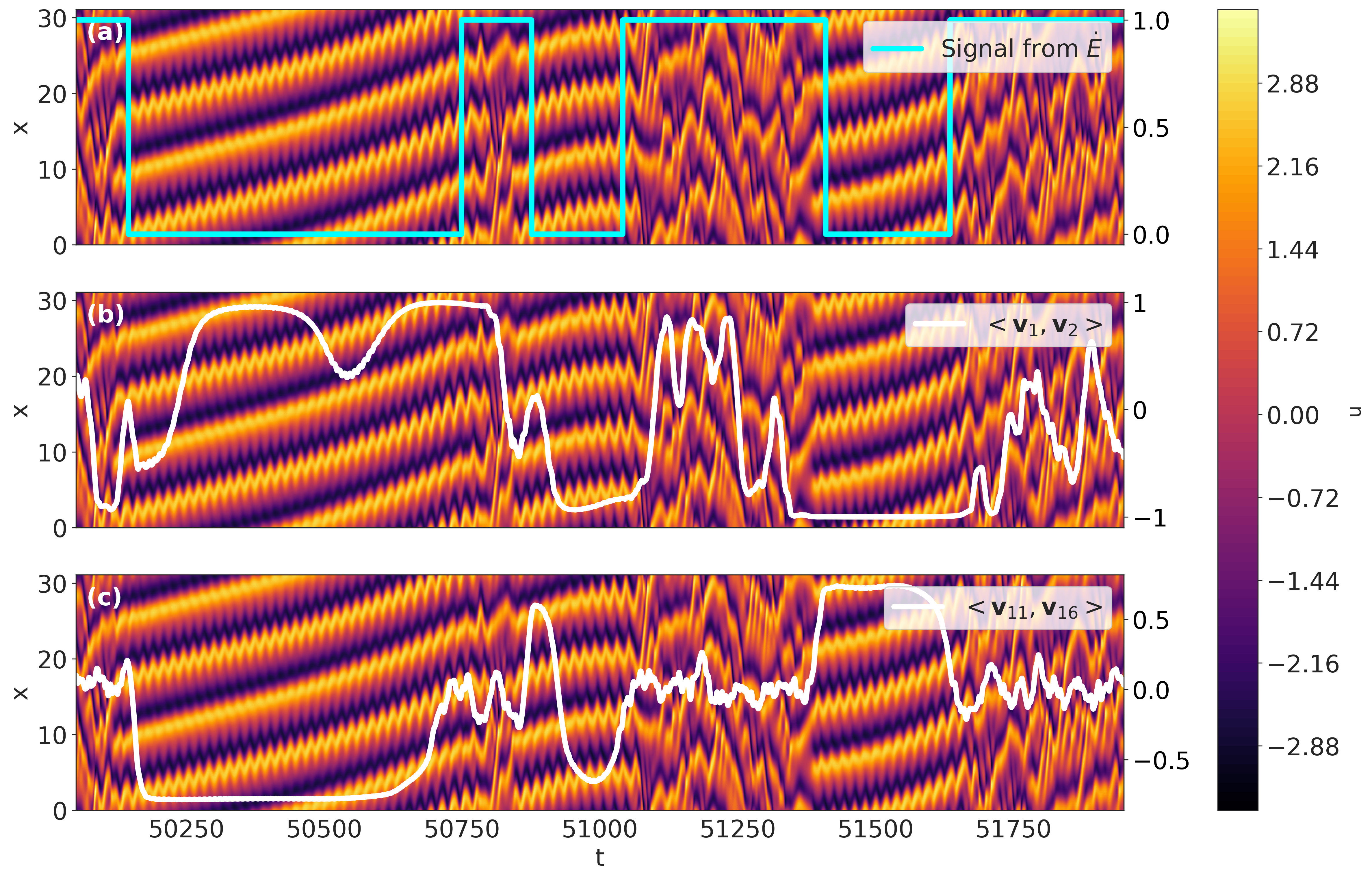}
    \caption{(a) Hovm{\"o}ller plot of the KSE and squared signal of the energy tendency, $\dot{\rm E}$ for classification of the burst; (b) time series of $\mathbf{v}_{21}$ and $\mathbf{v}_{22}$, (c) time series of $\mathbf{v}_{1}$ and $\mathbf{v}_{2}$}.       
    \label{fig:hovmoller}
\end{figure*}

The intermittent behavior of the KSE is particularly intriguing, characterized by the alternation between laminar and chaotic regimes. Moreover, as illustrated in Fig.~\ref{fig:hovmoller}, these regimes are not local but instead involve the entire spatial domain simultaneously. The absence of spatial inhomogeneities is attributable to the spatial translational symmetry of the system with periodic boundary conditions.  

The spatial homogeneity against transitions potentially simplifies the classification and statistical study of the KSE intermittency. In particular, by analyzing a single arbitrary site, one can extract information about the system's global behavior.  
Nevertheless, Fig.~\ref{fig:hovmoller} reveals that the propagation of periodic solutions is tilted relative to the temporal evolution of a single site. This misalignment hampers the use of a single time series to classify intermittency. On the other hand, the global (spatial) nature of the switching between regimes suggests employing a quantity that aggregates information over the entire spatial domain. We opt to use the time derivative of the energy, its tendency, $\dot{\rm E}$, with ${\rm E}(t) = ||u(x,t)||_2 = \sqrt{\int_{0}^{L}u^2(x,t){\rm d}x}$. 
As we shall see, the energy tendency is a powerful indicator of regimes' transition, with regions of high variability (large values of $\dot{\rm E}$) associated with chaotic bursts, and vice-versa for quasi-stationary, laminar phases.

We compute the time derivative of the energy, $\dot{\rm E}$, and use it to construct a square signal that distinguishes between laminar and chaotic phases based on energy variability. Specifically, by applying a threshold to $\dot{\rm E}$, we assign the value $+1$ to intervals where the energy fluctuations exceed the threshold (corresponding to chaotic bursts) and the value $0$ to intervals of low variability, which are identified as laminar phases. This binary signal allows for a clear segmentation of the dynamics, from which we compute the laminar lengths. The square-wave signal, overlaid on the Hovmöller diagram of the KSE solution in Fig.~\ref{fig:hovmoller}a, provides a remarkably clear identification of the distinct phases along the trajectory, successfully highlighting the alternation between laminar and chaotic dynamics.

\begin{figure}[H]
     \includegraphics[width=0.45\textwidth]{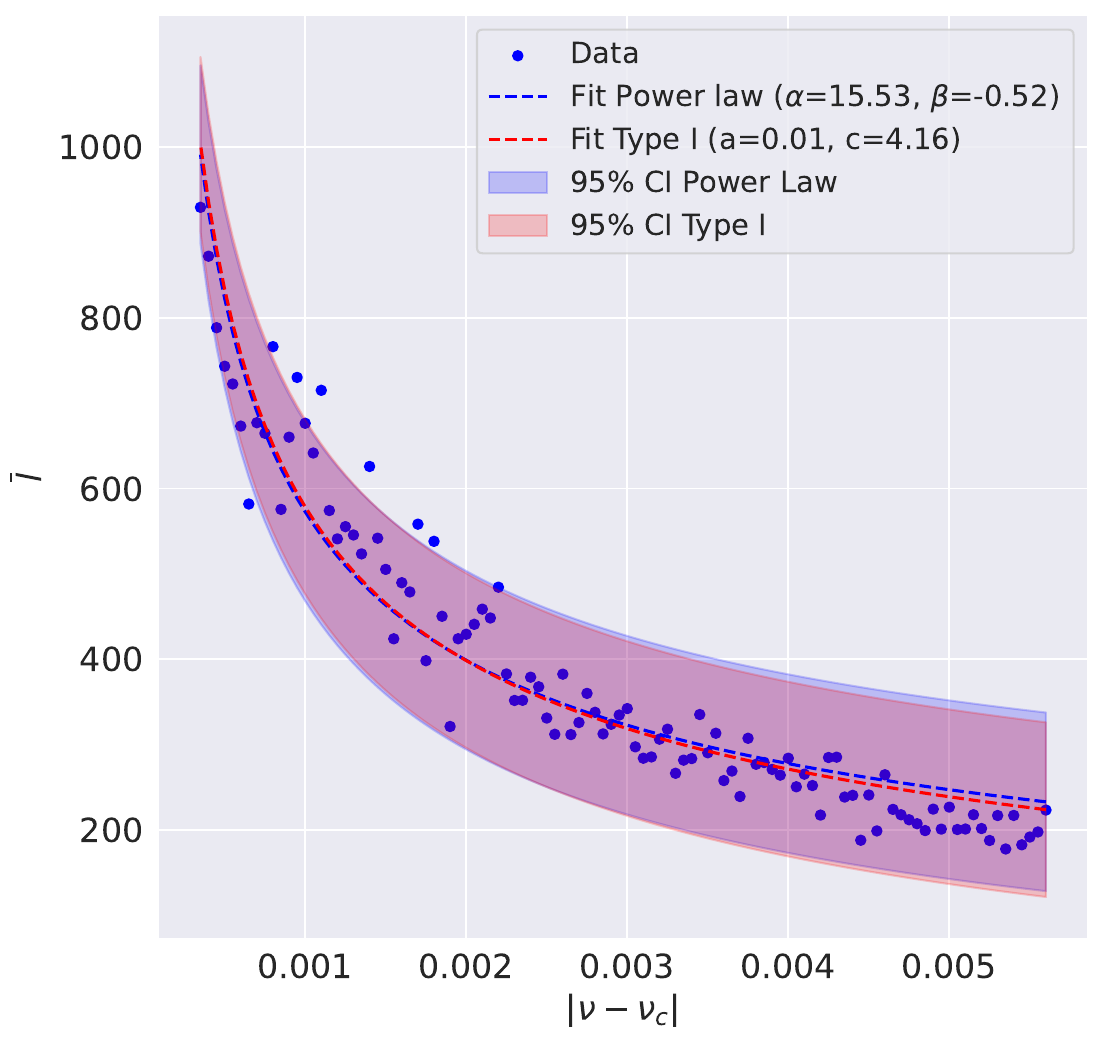}
    \caption{Numerically computed average laminar lengths $\overline{l}$ as a function of viscosity are shown as blue dots. The blue dashed line and the red dashed line represent, respectively, a type-I intermittency fit and a power-law scaling (see Table~\ref{tab:bifurcation_table}). The shaded areas around each line denote the corresponding $95\%$ confidence intervals.}
    \label{fig:fitKSE}
\end{figure}

From the squared energy tendency, we compute the average laminar length, $\overline{l}$, for various values of $\nu$ within the intermittent range. Results are shown in Fig.~\ref{fig:fitKSE} against deviation from the critical value, $\nu_c=0.8654$.

We fitted the data with Type-I scaling law, $ \overline{l} = \frac{1}{\sqrt{a \varepsilon}} arctan\left(c \sqrt{\frac{a}{\varepsilon}}\right)$ (see Sec.~\ref{sec:Intro} and Tab.~\ref{tab:bifurcation_table}) (red), whereby the celebrated parameters are $a$ and $c$. We then superimpose a power law fit of the form $f(x) = \alpha x ^\beta$, with the goal of evaluating the agreement with the characteristic relation, obtained in the limit $c\sqrt{a/\varepsilon}>>1$, {\it i.e.} when $\nu\rightarrow\nu_c$ (blue) (cf Sec.~\ref{sec:Intro}). 

The two fits exhibit strong agreement with data near the bifurcation threshold and begin to deviate only farther away, when the limiting characteristic relation loses its validity. Here, both the fits overestimate the value of $l$.  In particular, the power-law exponent closely matches the theoretical value $\overline{l} \propto |\nu-\nu_c|^{-1/2}$. The little deviation from the theoretical prediction is likely attributable to computational constraints. Near the bifurcation threshold, the laminar phases grow so long that obtaining a statistically meaningful number of samples for averaging becomes challenging.

\subsubsection{Degeneracy of LEs and continuous family of orbits}

The spectra of LEs are highly sensitive to the viscosity parameter $\nu$. Nevertheless, 
for all viscosity values considered here, the LEs are degenerate in the neutral portion, with two zero LEs. For $\nu=0.87$ the KSE displays a limit cycle with the two largest LEs being zero. These LEs remain equal to zero even when the viscosity is increased, for $\nu>0.87$. At first glance, a Lyapunov spectrum of the form $(0,0,-,-,-,\ldots)$ suggests that the system evolves on a torus~\citep{chaosmesure}. We shall see that this interpretation is inconsistent with our observations. 

To show this, we generated $20$ distinct initial conditions, distinct among them for perturbations of magnitude $\eta = 10 \cdot k$, with $k = 1, 2, \ldots, 20$. The initial energy injected in KSE depends on the initial conditions so that $E_0^{(k)}=||u^{(k)}_0||_2$. From these $20$ initial conditions we run $20$ simulations from where we observe the emergence of $K=20$ distinct periodic orbits, each associated with a different initial energy. The existence of this family of orbits is attributed to the symmetries in the KSE \citep{KSCvi, KS_POs1, KS_POs2}, like the spatial translational symmetry induced by the periodic boundary conditions, and the degeneracy of the two zero LEs in the presence of a periodic orbit is also induced by the symmetric property of the system. In this configuration, the initial condition acts effectively as a third control parameter - alongside $L$ and $\nu$ — determining the system’s evolution. For values of $\nu$ near the bifurcation point ($\nu \approx 0.865$), the KSE displays both the periodic orbits at low initial energy and strange attractors at higher initial energy levels.

The roughness of the LEs is most probably associated with the dependencies of the attractors on the initial state. This aspect will be further investigated and would indeed require experiment-dependent adjustments to the initial conditions to ensure the LEs' spectra are consistently reflecting parameter-driven modifications in the phase space. 
Moreover, as the initial energy increases, the period of the orbit decreases approximately as $T \propto 1/E_0$, while the amplitude, proportional to time-averaged energy, increases linearly with $E_0$.

\subsubsection{Local analysis}

We proceed with the local analysis fixing the parameter at $\nu=0.86$, with initial state $u_0$ obtained by populating the vector with random initial conditions, sampled from a uniform distribution in the interval $[0,1]$. For this configuration the LEs spectrum takes the form  $(+,+,+,+,0,0,-,...,-)$. With the splitting of the tangent space in mind (cf Sec.~\ref{sec:CLVs}), we conveniently consider the joint stable and the neutral subspaces: the unstable-neutral subspace $\mathcal{W}^+_0$, of dimension $dim(\mathcal{W}^+_0)=n_u=6$. Similarly we indicate the stable subspace as $\mathcal{W^-}$, with $dim(\mathcal{W^-}) = n-n_u= n_s=95$.

\begin{figure}[H]
\includegraphics[width=0.45\textwidth]{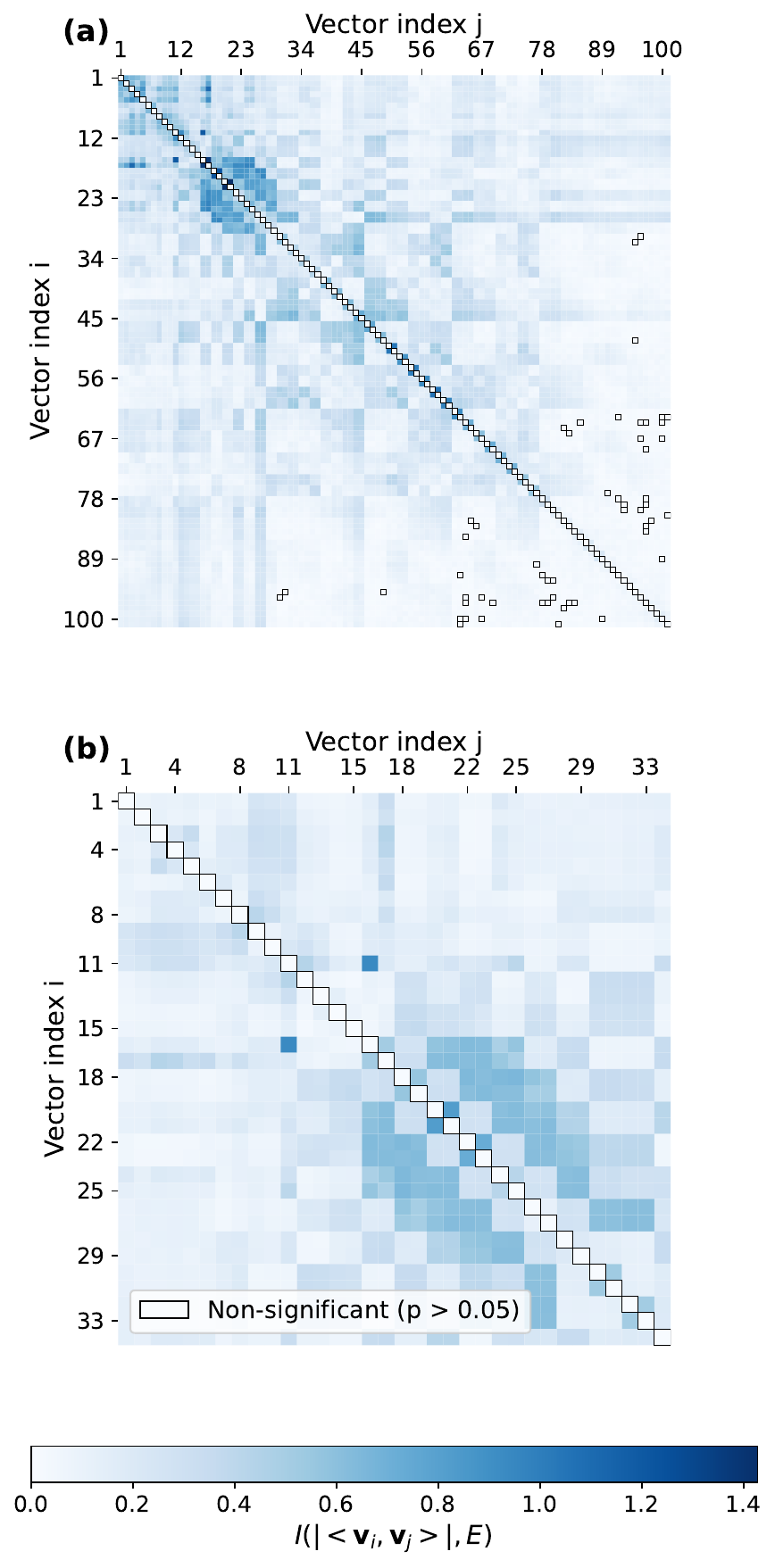}
    \caption{Mutual information between every angle combination of the CLVs and the energy of the system. The black squares mask the point where the mutual information is not significant.}
    \label{fig:CLVscorr}
\end{figure}

In Fig.~\ref{fig:CLVscorr}(a), we show the mutual information, computed with the k-NN method \cite{Kraskov2004}, between the angles of every pairs of CLVs and the energy, $I(|<\mathbf{v}_i,\mathbf{v}_j>|,E)$, computed using timeseries of 20,000 time steps. This is an information measure which quantifies the amount of dependence between two variables: it is zero if and only if the variables are statistically independent, and it increases as their statistical dependence increases, regardless of whether it is linear or nonlinear. A bootstrap-based hypothesis has been performed to assess the statistical independence between variables: non-significant values (p-value>0.05) are indicated by black squares. In these cases, mutual information does not provide a robust statistical measure. Overall, the results show that up to about one-third of the angles between pairs of CLVs share significant mutual information with the energy. Mutual information generally reduces with increasing vector index.

The CLVs (their angles thereof) between approximately the 15th and the 33rd have the largest values of mutual information. We therefore repeated the analysis on a longer timeseries of 200,000 timesteps, for solely the first 33 CLVs. Results are given in Fig.~\ref{fig:CLVscorr}(b) and show that the highest mutual information is observed for the vectors $|<\mathbf{v}_{11},\mathbf{v}_{16}>|$. This couple of vectors belong to the same stable subspace, yet they convey great information about the regime change. 

\begin{figure}[H]
\centering
\includegraphics[width=0.4\textwidth]{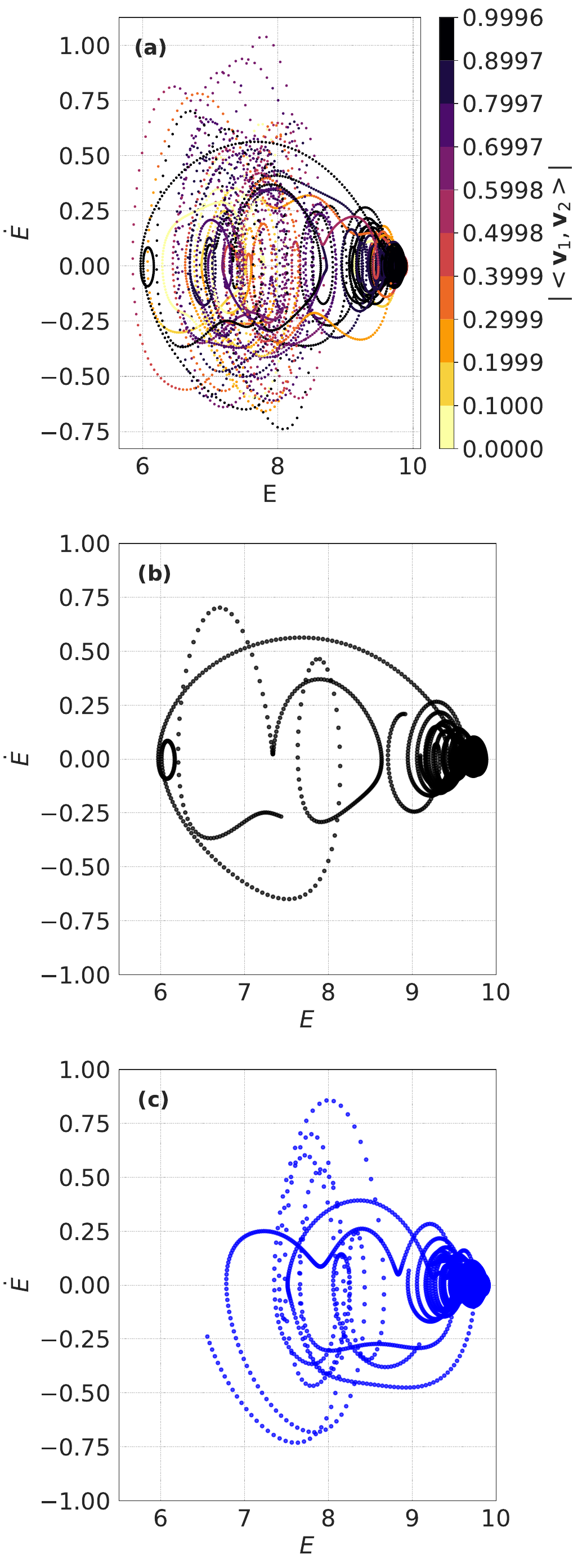}
    \caption{Energy space with mean cosine angle of the first two vectors (a), laminar region reconstructed from the CLVs  (b) and laminar region reconstructed from the energy signal (c).}
    \label{fig:ENspace}
\end{figure}

We believe that this is directly related to the framework proposed by \citet{Takeuchi-Ginelli2018}, which conjectured a decomposition of the KSE tangent space into physical and spurious modes. Following this line of arguments, the CLVs beyond the 33rd rank correspond to spurious modes that do not carry meaningful information about the system’s dynamics, thus neither about its intermittent behavior. This reasoning also helps to explain why highest mutual information' values arise by taking pairs of stable CLVs: despite being stable, and thus bounded to be damped asymptotically, 
they are physically consistent and thus potentially informative.
Finally, it is worth emphasizing the high informational content found in nearby or adjacent pairs of CLVs (those close to the diagonal). 

As stated in Sec.~\ref{sec:KSEglobal}, we can use the energy tendency to identify the chaotic bursts. These are indeed associated with a huge variability of the energy. By looking at trajectories in the energy space $(\dot{E}, E)$, we can identify a region of lower variability corresponding to the laminar phase. 
Similarly to what we showed in Sec.~\ref{sec:Results} for L63 for the identification of the limit cycle, in the KSE we are able to identify laminar and chaotic phases by looking if the angle between $\mathbf{v}_1$ and $\mathbf{v}_2$ are collinear. In Fig.~\ref{fig:ENspace}(a) we plot the system trajectory in the space $(\dot{{\rm E}}, {\rm E})$
and color it with the value of the angle between $\mathbf{v}_1$ and $\mathbf{v}_2$. A visual inspection already reveals that most of the portion of the trajectory corresponding to the laminar phase is characterized by high values of $|<\mathbf{v}_1, \mathbf{v}_2>|$, meaning strong alignment (collinearity) of the two CVLs occurs there, while for the burst it is the opposite scenario, with the vectors tending to be orthogonal to each other. 
We use this information to reconstruct the laminar region of the attractor. We use the signal from $|<\mathbf{v}_1, \mathbf{v}_2>|$, and sample the points in the energy space when $|<\mathbf{v}_1, \mathbf{v}_2>| \approx 0$ (Fig.~\ref{fig:ENspace}(b)) and compare them to the ones obtained by sampling the points corresponding to the "0" states of the squared signal obtained in Sec.~\ref{sec:KSEglobal} using a threshold on $\mathrm{\dot{E}}$ (Fig.~\ref{fig:ENspace}(c)).

The signal derived from energy is used as a target (Fig.~\ref{fig:ENspace}(c)). Indeed, our underlying hypothesis — see the success of $\dot{\mathrm{E}}$ in classifying the transition from laminar to chaotic phases (Fig.~\ref{fig:hovmoller})— is that energy is a reliable quantity capable of decomposing the energy space into two distinct regions: the laminar and the chaotic zones. 

It should also be emphasized that, when observing the energy space for a fully periodic setup, such as $\nu = 0.87$, all the energy would be confined within the very dense subspace containing values between $E \in (9, 10)$ and $\dot{E} \in (-0.25, 0.25)$.

These discrepancies between a fully laminar regime and the laminar regions in an intermittent regime are due to changes in the spatial characteristics of the periodic structures preceding the chaotic burst (see Fig.~\ref{fig:hovmoller}). This leads to a decrease in energy and an increase in variability, which is visible as larger amplitude in $\dot{E}$ (Fig.~\ref{fig:ENspace}).

We can therefore observe that the signal from the CLVs perfectly reproduces the very dense, high-energy, and low-variability region in which the non-intermittent periodic dynamics would be confined, thus highlighting that the CLVs are able to capture laminar structures. However, in the low-energy and high-variability region, there are discrepancies between the two indicators, which highlight different trajectories. From a statistical point of view, this minimal discrepancy is negligible considering the high density of points in the fully laminar zone. Indeed, to provide a quantitative measure of this result, we compute the Wasserstein distance, yielding a value of $WD=0.3$, highlighting a small difference.

The optimal pair of CLVs is expected to be model dependent, let alone that its identification can be difficult and computationally expensive. It is therefore desirable to find a general indicator with satisfactory performance and affordable computational cost. The unstable-neutral/stable subspaces splitting can play this role, whereby the angle between the subspace brings the signature of the regime changes. Following \citet{Bocquet_Carrassi_2017}, at each time step, we compute the angle between the unstable-neutral subspace $\mathcal{W}^+_0$ and a progressively larger portion of the stable subspace $\mathcal{W^-}$

\begin{equation}
    \Theta^{n_s}(t) =\sum_{j=n_{u}+1}^{n_s} \sum_{i=1}^{n_u} <\mathbf{v}_i(t),\mathbf{v}_j(t)>^2,
\end{equation}
with $n_u=6$ and $n_s=7, ..., 101$; recall that $n=101$.  

We therefore compute the mutual information between this indicator and the energy of the system. We first apply a rolling mean to both signals, using a right-aligned window of $100$ time steps. Then we study how $I(\Theta^{n_s}(t), E(t+\tau))$ varies as a function of the number of vectors included in the stable subspace ($n_s$) and the lag ($\tau$). This approach involves sequentially adding vectors from $\mathcal{W^-}$, starting with those corresponding to smaller amplitude LEs. The rationale behind this procedure lies in the fact that the stable vectors associated with smaller amplitude LEs carry the most significant amount of information.
With this analysis, we aim to determine whether a smaller number of vectors is sufficient to characterize the intermittency locally. 

\begin{figure}[H]
    \includegraphics[width=0.5\textwidth]{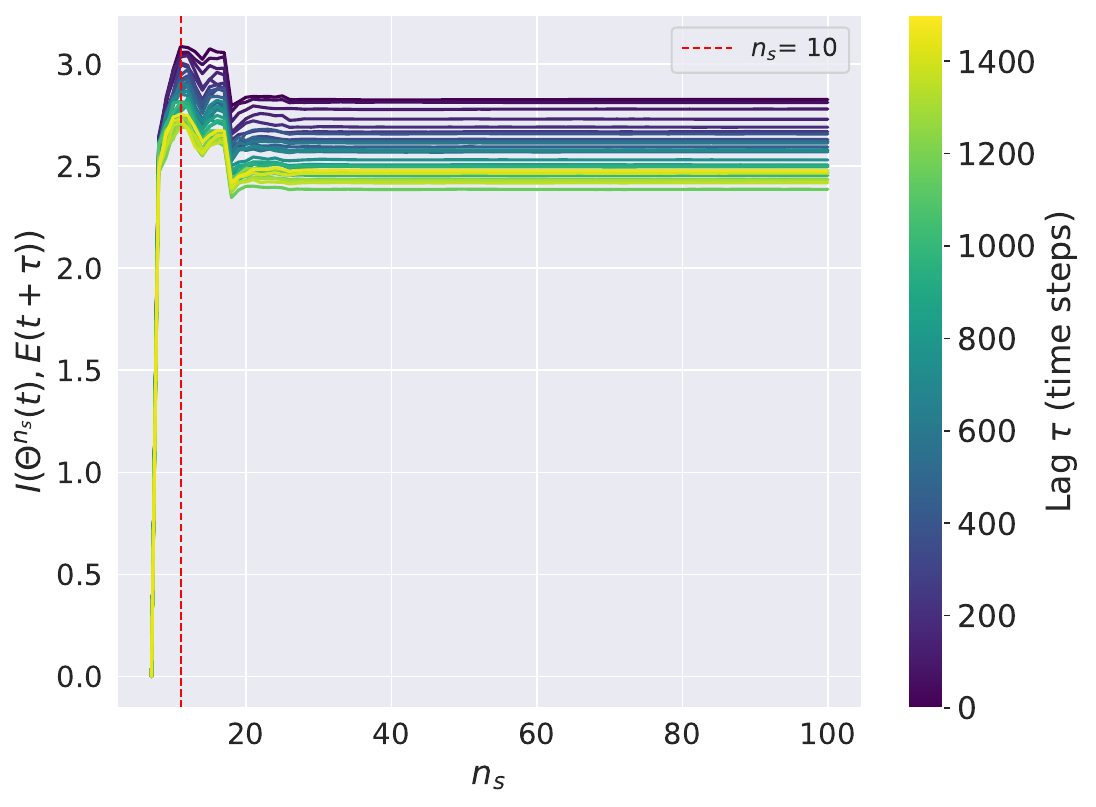}
    \caption{Mutual information between the mean separation of the two subspaces and the energy, as function of the number of vectors included and the lag. The red vertical line indicate the number of vectors for which we obtain maximum correlation and with the blue the maximum correlation for the single couple of vectors.}
    \label{fig:corrlag}
\end{figure}

In Fig.~\ref{fig:corrlag} we show the results of this analysis. Notably, with a zero time lag, the mutual information reaches its maximum when only $10$ stable CLVs are considered. Beyond this point, the correlation decreases and eventually saturates once more than 20 vectors are included. This behavior reinforces our hypothesis about the importance of the splitting discussed by \citet{Takeuchi-Ginelli2018}, confirming that adding vectors with indices greater than 20 does not contribute any meaningful insight into the characterization of the intermittent regime.

We also observe that mutual information decreases as the lag increases. However, this decline is rather slow, suggesting that a high level of information is retained across a range of lag values. This highlights not only the indicative value of the mutual information but also its potential predictive power.

To conclude, we emphasize once again that the key result of this analysis is the substantial reduction in the dimension of the driver indicators of the regime shift: we have reduced it from an initial 101 dimensions to just 16, effectively capturing its behavior with significantly fewer vectors. This reduction not only enables a deeper physical interpretation of the system — particularly in terms of how the geometry of the tangent space reflects the intermittent dynamics — but also provides a practical tool for investigating such transitions using a reduced set of physically meaningful observables, thereby lowering the overall computational cost.

\section{\label{sec:concl} Conclusion}
By analyzing type-I intermittency in the L63 system, we observed that the LEs and the mean along the flow of the angle between the CLVs globally identify the bifurcation point where the limit cycle transitions into a strange attractor. We also found that the expansion of the attractor within the state space, driven by the increasing duration of chaotic bursts, can be quantified by observing the fractal dimension, which follows a power law. Specifically, for type-I intermittency, we demonstrated that this growth can be computed near the bifurcation point using the scaling relationship of the laminar length, $d \propto \overline{l}^{-2 \gamma}$. Furthermore, we showed that the angles between CLVs performed exceptionally well in detecting the local transition from one regime to another. Then we used the angle between the first and the neutral vector to reproduce the structure of the limit cycle. 

We then move on to study merging crises intermittency. While the angles between CLVs proved effective in detecting the transitions associated with crisis-induced intermittency in the double-well potential system, they failed to do so in the L96 model. In the latter case, we emphasized the distinctive features of the bifurcation diagram, which reveals a wide variety of dynamical behaviors as the forcing parameter varies within its negative range. In particular, we identified the occurrence of a merging crisis, with both attractors possessing periodic windows. We further demonstrated that these oscillations may manifest as laminar episodes within the intermittent time series generated by the merging crisis. Interestingly, we found that the last two CLVs in the L96 model were able to capture these laminar intermissions with remarkable accuracy.

We then investigated on-off intermittency. Here, we observed that every combination of CLVs provided information to indicate regime transitions. For certain combinations, not only they act as indicators, but they also exhibit predictive potential by anticipating chaotic bursts. Lastly, we noted that whenever the last vector was included, the time series of the angle oscillated during the ``off'' state. A similar behavior was observed in L96 during the periodic intermissions.

Next, we discovered an intermittent window in the Kuramoto–Sivashinsky equation. Here, the transition occurred globally across the entire spatial domain, prompting us to use energy to characterize the system globally. From this, we calculated the mean laminar length, whose scaling law appears consistent with that of type-I intermittency. After the bifurcation, where the system transitions from a strange attractor to a periodic orbit, we observed that the spectrum of LEs exhibited the form $(0,0,-,-,-,...,-)$, which might suggest that the system evolves on a torus. However, this interpretation contradicts the observations. We showed that the degeneracy of the two zero exponents arises from the system's translational symmetry, implying the existence of a family of orbits for the same parameter value. Moreover, we showed that the amplitude of these orbits grows linearly with the initial energy, while the period decreases following $T \propto 1/E_{0}$. 

When doing the local analysis in the KSE, we observed that various combination of the CLVs provided information about the regime change. Specifically, we conducted a study on the mutual information between each combination of CLVs and the system's energy. We found that $I(|<\mathbf{v}_i,\mathbf{v}_j>|,E)$ tended to decrease for higher-index vectors. We believe that this decay in information when moving towards more stable CLVs can be understood through the lens of the splitting introduced by \citet{Takeuchi-Ginelli2018}, providing a rationale for the existence of a subgroup of ``physical'' CLVs that effectively capture and indicate the intermittent regime. This interpretation also accounts for the fact that the highest mutual information was observed between the 11th and 16th CLVs, which — despite both belonging to the stable subspace — may still convey meaningful information as they might belong to the subgroup of physical modes.

Furthermore, we found that the alignment of the CLVs can be used to pinpoint a specific region of the energy space that corresponds to the laminar phase, analogous to the approach used to reconstruct the reminiscence of the limit cycle in L63.

Additionally, we developed a global indicator $\Theta$ that measures the mean separation between the unstable-neutral and stable subspaces. We analyzed the mutual information between $\Theta$ and the energy. This was done by varying the 
number of stable CLVs included in the computation of the global angle with the unstable-neutral subspace. By doing so, we showed that the maximum of $I(\Theta^{n_s}(t), E(t+\tau))$ was obtained by considering the first 10 vectors and that it saturated after 20 vectors. This further confirms that including vectors with indices above 20 does not add significant information for characterizing the intermittent regime.

In conclusion, the methodologies employed in this study highlight a considerable degree of generality in the use of LEs and CLVs for the classification and prediction of regime transitions across different intermittent systems. In particular, we believe that CLVs can serve as powerful tools for gaining deeper insight into the local dynamics of intermittent systems. They offer the potential to uncover key dynamical features, to indicate - and even predict - transitions between intermittent regimes, and to operate with a significantly reduced set of quantities compared to the full system dimension.
These aspects will be addressed in the future together with their application to higher complexity systems. 

\section*{Supplementary material}

Supplementary material provides additional information on the scaling behavior of crisis-induced intermittency near the critical forcing value. It includes a figure (Fig.~S1) showing the average burst time as a function of the distance from the critical parameter value, along with a linear fit in log-log scale. This supports the theoretical discussion and the result presented in Table~\ref{tab:bifurcation_table}.

\section*{Data Availability Statement}

The data that support the findings of this study are available from the corresponding author upon reasonable request. 

\section*{Acknowledgments}
AB and AC have been funded as members of the Scale-Aware Sea Ice Project (SASIP) supported by grant G-24-66154 of Schmidt Sciences, LLC –– a philanthropy that propels scientific knowledge and breakthroughs towards a thriving world.

\section*{Author declarations}

\subsection{Conflict of Interest}

The authors have no conflicts to disclose.

\subsection{Author Contributions}

{\bf Alessandro Barone}: Conceptualization (lead); Formal analysis (lead); Investigation (lead); Methodology (lead); Software (lead); Validation (lead); Visualization (lead); Writing– original draft (lead); Writing– review \& editing (lead).  {\bf Alberto Carrassi}: Conceptualization (supporting); Investigation (supporting); Formal analysis (supporting); Methodology (supporting);
Validation (supporting); Writing– original draft (supporting); Writing– review \& editing (supporting). {\bf Thomas Savary}: Conceptualization (supporting); Software (supporting); Investigation (supporting); Methodology (supporting). {\bf Jonathan Demaeyer}: Conceptualization (supporting); Investigation (supporting); Methodology (supporting);
Validation (supporting); Writing– review \& editing (supporting). {\bf Stephane Vannitsem}: Conceptualization (supporting); Investigation (supporting); Methodology (supporting);
Validation (supporting); Writing– review \& editing (supporting).

\section*{Preprint notice}

\noindent\textcopyright\ (2025) A. Barone, A. Carrassi, T. Savary, J. Demaeyer, S. Vannitsem. 
This article is distributed under a Creative Commons Attribution-NonCommercial-NoDerivs 4.0 International (CC BY-NC-ND) License. 
\url{https://creativecommons.org/licenses/by-nc-nd/4.0/}

\section*{references}

\end{document}